%% file: main.tex
\documentclass[10pt,twocolumn]{article} 
\usepackage{arxiv}
\usepackage{graphicx}
\usepackage{epsfig}
\usepackage{subcaption}
\usepackage{enumitem}

\AtBeginDocument{%
  \providecommand\BibTeX{{%
    \normalfont B\kern-0.5em{\scshape i\kern-0.25em b}\kern-0.8em\TeX}}}

\newcommand{\name}{Proteus} 
\usepackage[normalem]{ulem}
\usepackage{epsfig}
\usepackage{color}
\usepackage{graphicx}
\usepackage{amsmath}
\usepackage{url}
\usepackage{mwe}
\usepackage{epsfig,endnotes}
\usepackage{subcaption}
\usepackage{array}
\usepackage{tabularx} 
\usepackage{grffile}

\begin{document}
\title{No Size Fits All: Automated Radio Configuration for LPWANs \newline}

%\author{Deepak Vasisht}
%\email{deepakv@illinois.edu}
%\affiliation{
%  \institution{Department of Computer Science and Engineering, University of Illinois} %\city{Urbana}
%  \state{IL}}
%\author{Tusher Chakraborty}
%\email{Tusher.Chakraborty@microsoft.com}
%\affiliation{%
%  \institution{Microsoft}
%  \city{Redmond}
%  \state{WA}
%}
%\author{Joshua R. Smith}
%\email{jrs@cs.washington.edu}
%\affiliation{%
%  \institution{Department of Electrical and Computer Engineering and Computer Science and Engineering, University of Washington}
%  \city{Seattle}
%  \state{WA}
%}
%\author{Ranveer Chandra}
%\email{ranveer@microsoft.com}
%\affiliation{%
%  \institution{Microsoft}
%  \city{Redmond}
%  \state{WA}
%}

\author{
Zerina Kapetanovic\textsuperscript{1*},
Deepak Vasisht\textsuperscript{2},
Tusher Chakraborty\textsuperscript{3},
Joshua R. Smith\textsuperscript{1},
Ranveer Chandra\textsuperscript{3}
\\
\bf{1} University of Washington, Seattle, WA, \\
\bf{2} University of Illinois, Urbana, IL, \\
\bf{3} Microsoft, Redmond, WA \\
\bigskip
*zerinak@uw.edu
}
\maketitle 

\begin{abstract}
Low power long-range networks like LoRa have become increasingly mainstream for Internet of Things deployments. Given the versatility of applications that these protocols enable, they support many data rates and bandwidths. Yet, for a given network that supports hundreds of devices over multiple miles, the network operator typically needs to specify the same configuration or among a small subset of configurations for all the client devices to communicate with the gateway. This one-size-fits-all approach is highly inefficient in large networks. We propose an alternative approach -- we allow network devices to transmit at any data rate they choose. The gateway uses the first few symbols in the preamble to classify the correct data rate, switches its configuration, and then decodes the data. Our design leverages the inherent asymmetry in outdoor IoT deployments where the clients are power-starved and resource-constrained, but the gateway is not. Our gateway design, Proteus, runs a neural network architecture and is backward compatible with existing LoRa protocols. Our experiments reveal that Proteus can identify the correct configuration with over 97\% accuracy in both indoor and outdoor deployments. Our network architecture leads to a 3.8 to 11 times increase in throughput for our LoRa testbed.

\end{abstract}

%\subsection*{Abstract}
%\input{abstract}

\input{intro-v4.tex}

\input{lora_primer.tex}

\input{challenges.tex}

\input{method-3.tex}

\input{implementation.tex}

\input{results.tex}

\input{related.tex}
\input{conclusion.tex}

\section*{Dataset Availability}
We collected datasets with several hundred thousand LoRa symbols in multiple settings. These datasets will be available publicly to the community to develop new algorithms.

 %as described in our evaluation

\bibliographystyle{abbrv}
\bibliography{main.bib}

\end{document}

%% file: intro-v4.tex
\begin{figure}[h!]
\centering
\begin{subfigure}[b]{0.4\textwidth}\includegraphics[width=\textwidth]{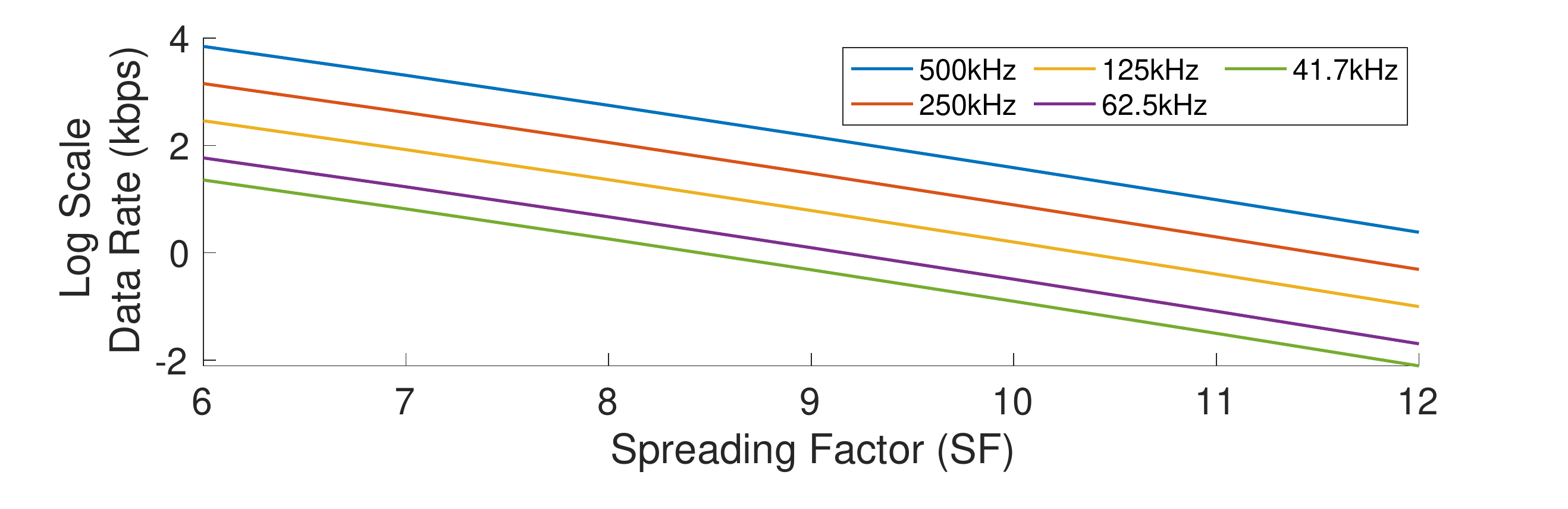}
\caption{LoRa Data Rate. \textmd{The data rate for LoRa decreases as a function of spreading factor and increases with bandwidth.}}
\label{fig:lora_dr}
\end{subfigure}

\begin{subfigure}[b]{0.4\textwidth} \includegraphics[width=\textwidth]{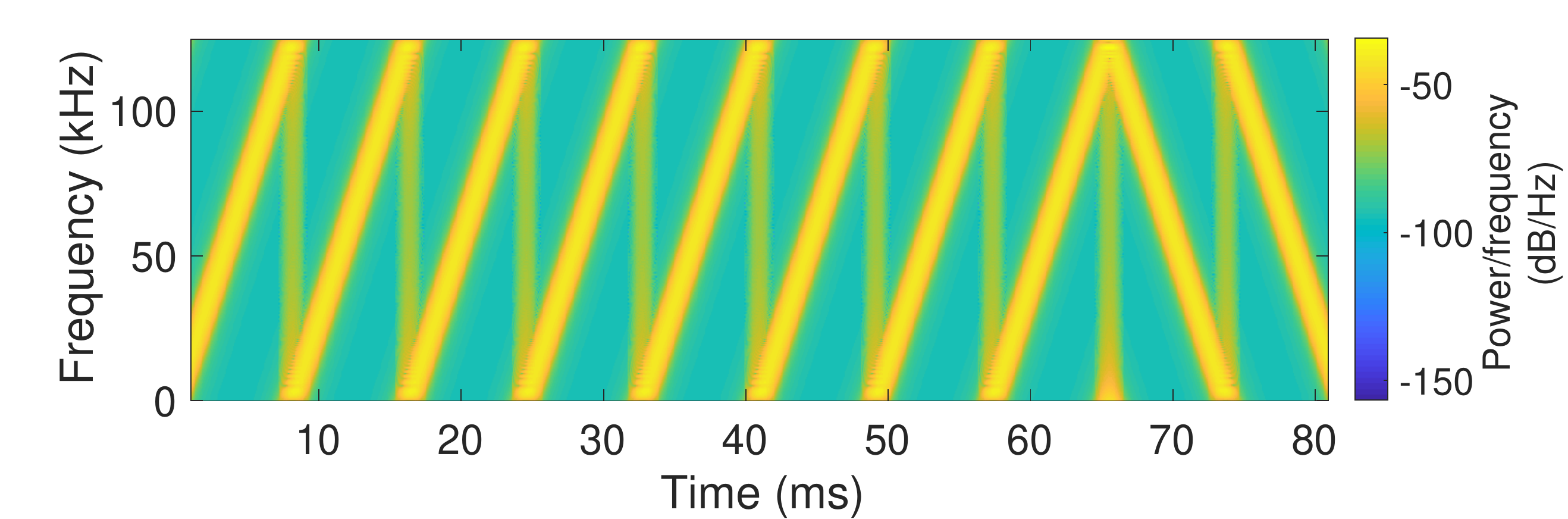}
\caption{LoRa Preamble. \textmd{The spectrogram of a LoRa preamble with 8 upchirps and 2 downchirps.}}
\label{fig:envelope}
\end{subfigure}
%\vspace{-0.2in}
\caption{Characteristics of the LoRa protocol.}

\label{fig:lora_pm}
\end{figure}

\section{Introduction}

Low Power Wide Area Networks (LPWANs), such as LoRaWAN, have become increasingly popular for large-scale Internet of Things (IoT) deployments. Despite their nascence, there are over 100 million devices using LoRaWAN already in deployment, with this number expected to exceed 730 million by 2023~\cite{iot_devices}. LPWANs can operate at lower power in comparison to other mainstream solutions, communicate over long distances, and are low-cost. These characteristics make such radios ideal for low throughput large-scale networks in cities, agriculture, forests, and many other industries. 

To support long range and diverse device requirements, LoRaWAN can operate at many different data rates. As shown in Fig.~\ref{fig:lora_dr}, the data rate is configured using two parameters: bandwidth of the chirp used in LoRa transmissions and the spreading factor~\footnote{The actual data rate also depends on the code rate used to ensure error correction. For this paper, we assume a fixed code rate}. As expected, higher bandwidth enables higher data rate. The spreading factor defines the time it takes to transmit one chirp, i.e. higher spreading factor means it takes longer to transmit a signal and hence lower data rate. The popular LoRa implementations can support bandwidths from 7.8 kHz to 500 kHz and spreading factors from 6 to 12 (on the log scale). Therefore, a device transmitting at 7.8 kHz and spreading factor 12 will achieve a {$\sim$1189} times less data rate than a device transmitting at 500 kHz and a spreading factor of 7.
%bb=0 0 100 100

In spite of this wide range of possibilities for devices in a network, the current paradigm requires a system designer to configure a single configuration setting of bandwidth and spreading factor for \textit{the entire network}, i.e., the bandwidth and the spreading factor is the same for \textit{all the devices}. Even though LoRaWAN Automatic Data Rate (ADR) algorithms have been proposed, they can take hours to days to converge~\cite{ADR_Issues, ADR_sim}, have significant control overhead, and don't handle multiple bandwidths. Such design choices stem from the need to limit complexity of the network and to reduce control overhead for low power IoT devices. However, this design choice also has multiple shortcomings.

First, this approach reduces the overall throughput of the system. A single LoRa gateway\footnote{LoRaWAN uses a gateway-client mode of operation.} is designed to cover a range over 10 Km with thousands of devices. In such large deployments, the devices at the end of range can barely support the lower data rates. As a result, this `single-size-fits-all' design forces even the devices that are closer to the gateway and can support high data rates to operate at an extremely low data rate. This reduces the overall network throughput and reduces the number of devices that it can support, by up to two orders of magnitude. 

Second, the `optimal' configuration for the gateway needs to be set by the network operator. Typically, this is done by trying out multiple configurations and choosing one that works for all client devices. Note that this process requires technical labor and is not always available; for instance, when deploying such devices in remote rural areas for agriculture monitoring. Moreover, the configuration selection needs to be dynamic. Due to changes in the environment or due to incremental deployment of devices, this configuration will stop working for a subset of the devices over time and will need frequent updates.

One might wonder why these networks do not leverage the extensive past work in rate adaptation protocols used in technologies like Wi-Fi, where the lowest data rate is used to send the preamble and coordinate the data rate configuration. Such rate adaptation is ill-suited to LPWANs because of three reasons. First, the overhead scales with the number of devices. Since LPWANs support hundreds of devices for a single gateway and each device transmits limited amount of data, this overhead becomes unsustainable. Second, the ratio of the highest rate to lowest rate is higher in LPWANs, therefore the preamble at lowest data rate will cause additional delays. Finally, the long preamble and the need to coordinate drains the limited battery of IoT devices deployed using LPWANs. Ideally, we want a rate adaptation system that has zero overhead for the clients and yet, flexible enough to accommodate multiple different data rates.

In this paper, we develop a new approach to rate adaptation that shifts all the burden of rate adaptation to the gateway. Our approach does not require the client devices and the gateway to agree on a rate a priori. Instead, each client transmits at a rate ideally suited to itself. A client far away may use a lower rate and a client close by may use a higher rate. The clients can choose the best-fit spreading factor and bandwidth. At the gateway, a neural network uses the first few symbols of the preamble to identify the data rate configuration being used by the client and tune the gateway to the right settings.

We exploit the preamble of a LoRa packet to enable our approach. Specifically, as shown in Fig.~\ref{fig:lora_pm}, the preamble consists of $N$ upchirps followed by two downchirps. The number of upchirps in the preamble is variable and can vary per packet per client. Therefore, if we can use the first one or two upchirps to identify the bandwidth and the spreading factor of the received chirps, we can tune the gateway to the right configuration. The gateway sees the rest of the preamble, which is in itself a valid preamble (and packet). Therefore, (a) our system is compatible with existing gateways (the gateways don't realize our system exists), and (b) the gateway doesn't need to maintain per-client state or coordinate with the client. In fact, the gateway doesn't even need to know which client the packet is from. The reconfiguration can happen at a per packet level.

\begin{figure}[t]
\vspace{0pt}
\centering
\includegraphics[width = 0.45\textwidth]{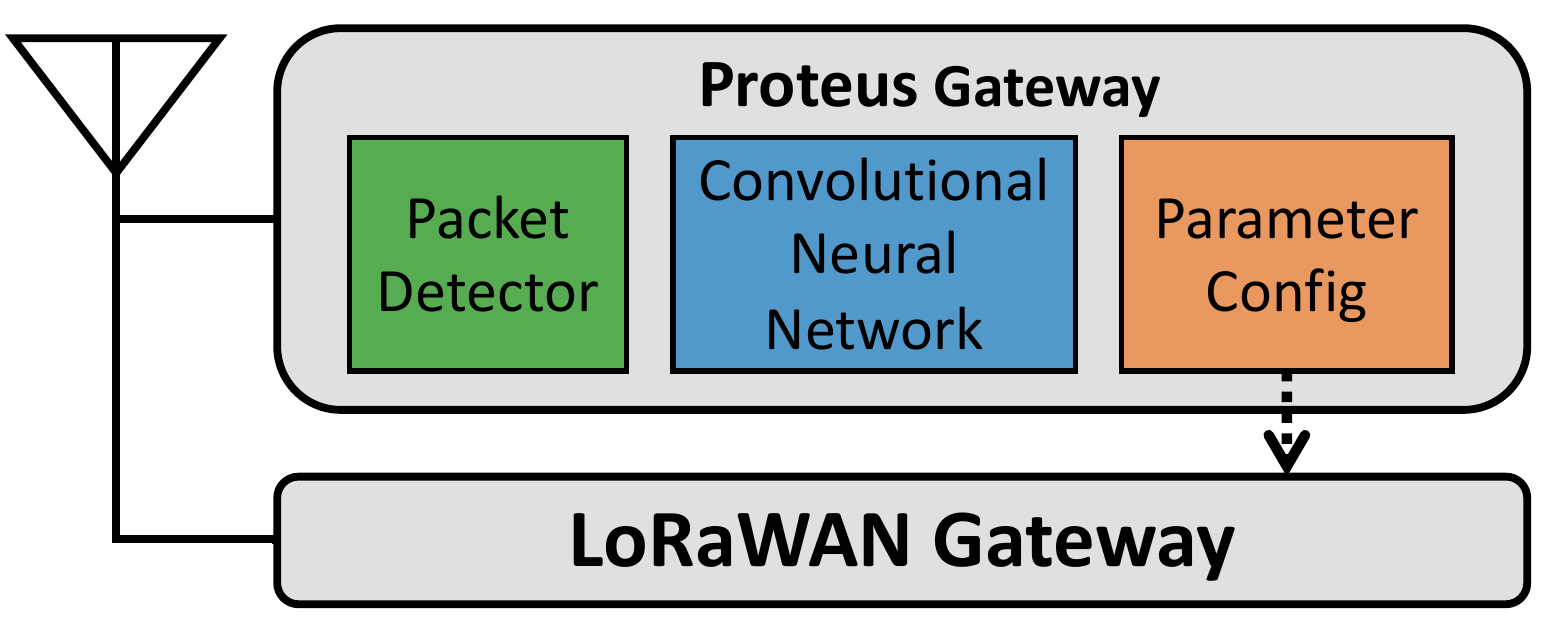}

\caption{Components of the Proteus System}
\vspace{-0.1in}
\label{fig:arch}
\end{figure}

We integrate this rate adaptation protocol in our gateway design, \name, as shown in Fig.~\ref{fig:arch}. In our current implementation, we use a Software-Defined Radio (SDR) in front of a standard LoRa receiver (future iterations could use newer LoRa hardware, e.g. SX1257~\cite{sx1257} that enable access to raw IQ samples off-the-shelf). The SDR detects the preamble, identifies the bandwidth and spreading factor of the signal from the first few upchirps, and quickly tunes the radio configuration of the LoRa radio to the right settings to receive the packet. This allows the gateway to successfully receive packets from clients operating at any configuration. Since this approach operates on a per-packet level, it supports data rate changes due to client mobility as well as dynamic changes in the environment.

For the {\name} design to work in the real world, we need to solve three key challenges (more details in Sec.~\ref{sec:challenges}):

\begin{itemize}
\item\textbf{Sensitivity:} To maintain LoRa's long-range, the system should be able to operate at very low SNRs. 

\item\textbf{Real-time Operation:} A packet must be detected by the SDR, and the LoRa radio should be reconfigured in real-time to ensure that the packet is not lost.

\item\textbf{Compatibility with existing deployments:} The system should not require protocol or client hardware changes to existing LoRa devices.
\end{itemize}

In, Sec.~\ref{sec:method}, we delve into how we solve these challenges using \name's neural network design and an innovative signal processing mechanism. We implement \name's gateway using the USRP SDR platform and evaluate it with off-the-shelf LoRa chipsets as clients. We evaluate \name\ in a broad variety of settings: benchtop experiments with varying signal strengths, an indoor deployment across multiple rooms, and an outdoor campus deployment. We summarize our results below:
\begin{itemize} [leftmargin=*, nosep]
    \item \name\ increases the throughput of an in-the-wild LoRa test bed by 3.8x to 11x using a single LoRa gateway.
    \item \name's configuration detection algorithm achieves a detection accuracy of 99.8\%, 95\%, and 98.2\% in indoors, outdoors, and benchtop experiments respectively. In contrast, an auto-correlation baseline achieves an accuracy of 67.4\%, 67\%, and 78\% respectively.
    \item \name's continues to operate effectively at low SNRs: achieves an accuracy of 94\% even when the signal is attenuated by over 140dB.
    \item \name's algorithm can generalize effectively to new environments and continues to operate in dynamic environments across time. In an experiment lasting 5 days, the accuracy of \name\ was consistently over 99\% with minor daily variations.

\end{itemize}

Finally, we believe \name\ opens up a new direction in rate adaptation algorithms. As neural networks evolve and get faster hardware implementations, can they allow the rate adaptation burden to be shifted to infrastructure alone and relieve the mobile devices from this overhead? This question would have implications beyond LPWANs. While we do not believe \name\ in the current form solves this broader question, we hope future iterations of \name\ would.

%% file: lora_primer.tex
\section{A Primer on LoRa}
\begin{figure}
    \centering	
    \begin{subfigure}[b]{0.22\textwidth}
    \includegraphics[width=\textwidth]{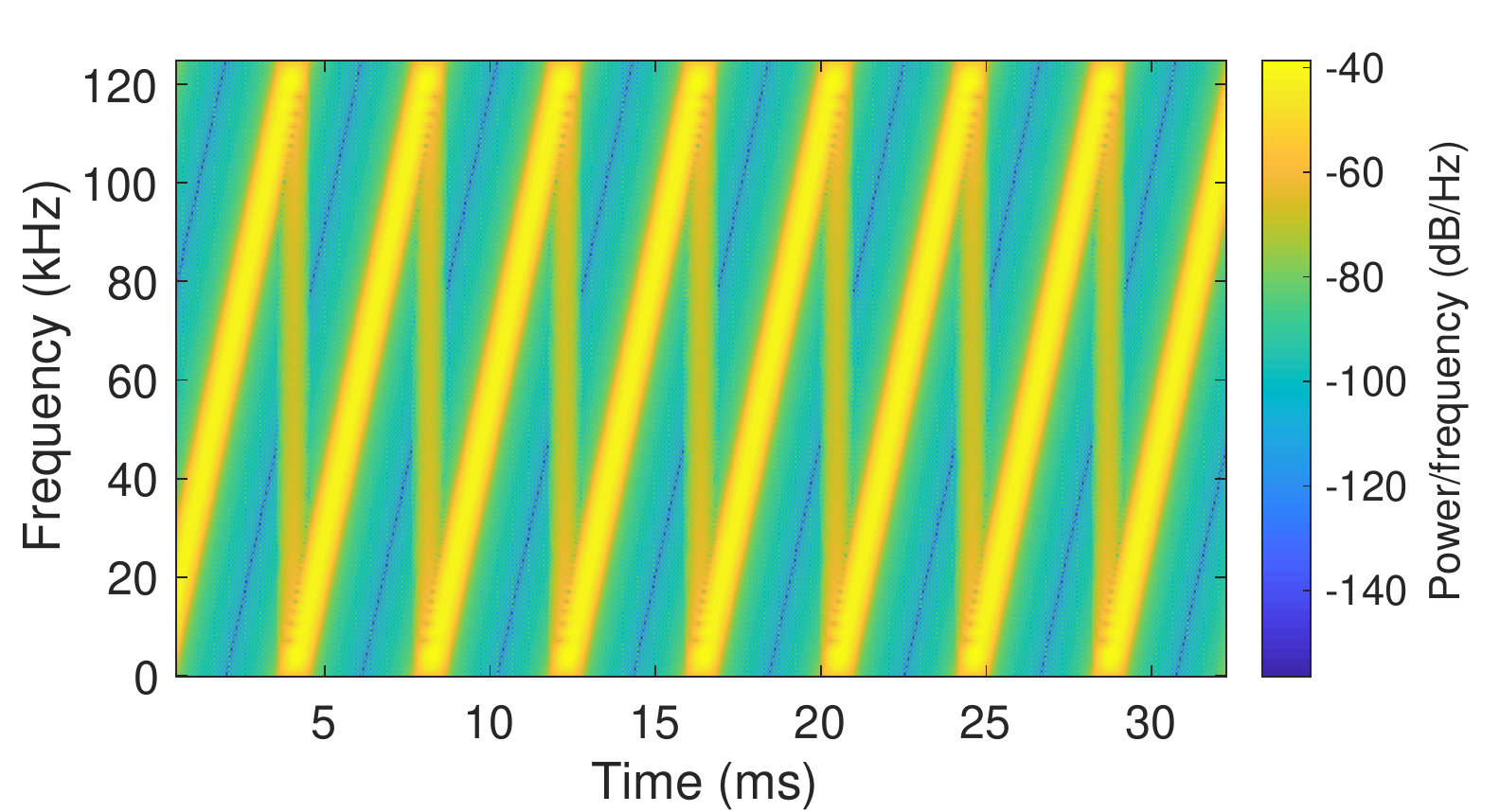}
    \caption{\textmd{BW=125kHz, SF=9}}
    \label{fig:chirp1}
    \end{subfigure}
    \quad
    \begin{subfigure}[b]{0.22\textwidth}
    \includegraphics[width=\textwidth]{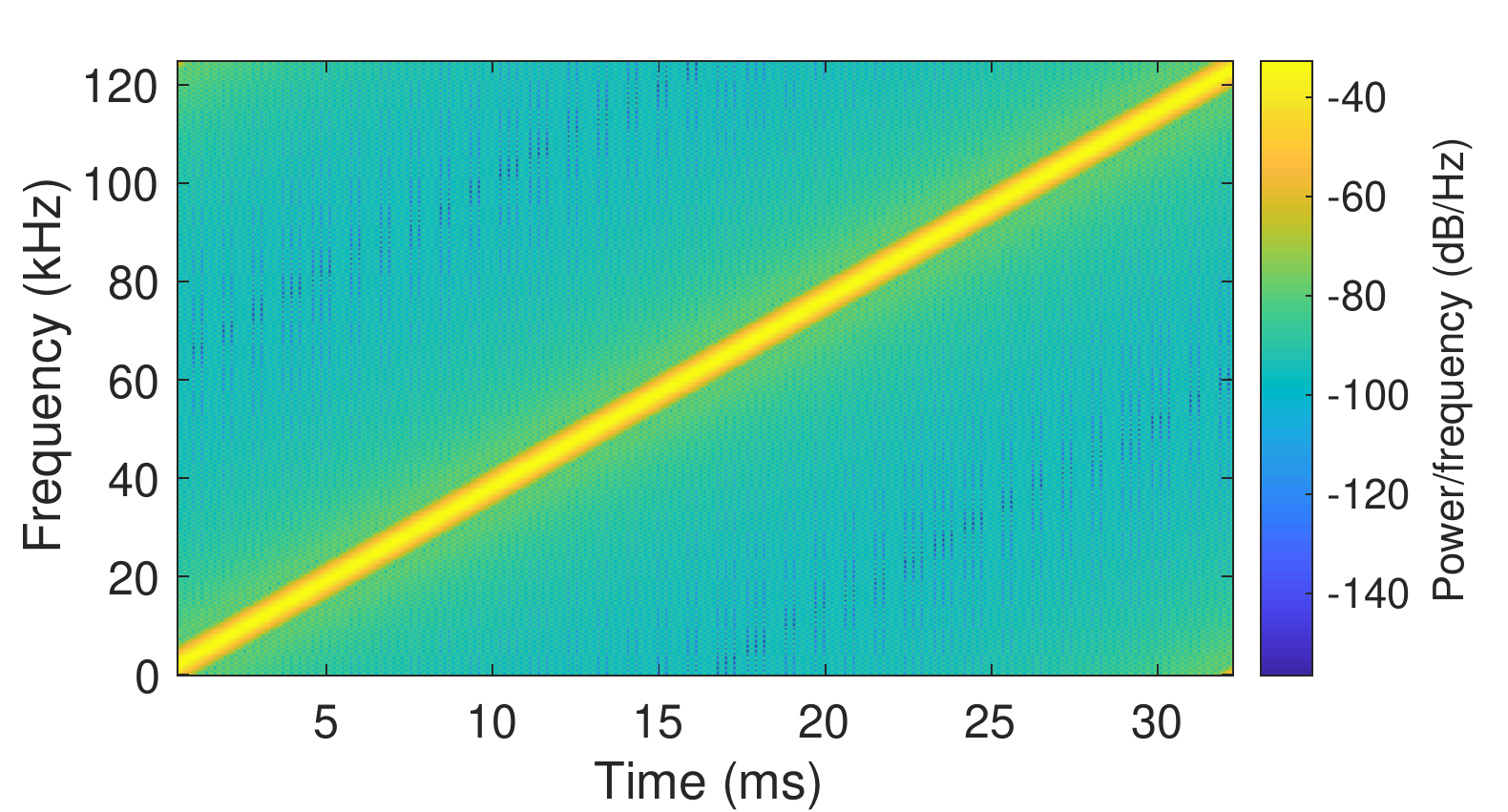}
    \caption{\textmd{BW=125kHz, SF=12}}
    \label{fig:chirp2}
    \end{subfigure}
    \quad
    \vspace{-0.2in}
    \caption{Spectrogram of LoRa Upchirps. \textmd{A comparison of chirps using different bandwidths and spreading factors.}}
    \vspace{-0.2in}
    \label{fig:chirps}
\end{figure}

LoRa is a physical layer implementation for LPWANs based on chirp spread spectrum (CSS) ~\cite{lora_mod}. In LoRa modulation, chirp signals are generated to encode data symbols. The frequency of the chirp varies linearly with time as shown in Fig.~\ref{fig:chirps}. Two parameters define the effective data rate: bandwidth and spreading factor. The BW controls the total span of the chirp in the frequency domain. The SF defines how long each chirp is in the time domain. Specifically, for a chirp with spreading factor $SF$, the time taken to transmit it is directly proportional to $2^{SF}$.

The time taken to transmit a chirp, $T_s$ is therefore given by, $T_s=\frac{2^{SF}}{BW}$, where $BW$ is the chirp and $SF$ is the spreading factor. Therefore, higher BW reduces the duration of each chirp and higher SF exponentially increases the duration of each chirp. 

To communicate data bits, the transmitter modifies the initial frequency of the chirp, $f$. Specifically, to send symbol value $S$, the transmitter sets the starting frequency to:
\begin{equation}\label{eqn:fs}\vspace{-0.1in}
    f(S) = S\times\frac{BW}{2^{SF}}
\end{equation}
LoRa allows $S$ to take values in the range $\{0,2,...,2^{SF}\}$. Thus, one chirp communicates one symbol which communicates $SF$ bits. As a result, the effective data rate, $R$, for a LoRa transmission is:
\begin{align}\label{eqn:rate}
    R &= SF\times\frac{1}{T_s} = SF\times\frac{BW}{2^{SF}}
\end{align}

As shown in Eq.~\ref{eqn:rate}, increase in BW increases the data rate. Decrease in SF increases the data rate. One might wonder if higher SF lowers the rate, why use it at all. This is because higher SF also increases the duration of a symbol making it easier to correctly decode. 

Finally, we reiterate the terminology for the rest of the paper. A symbol is a unit of data that is conveyed by each chirp. The duration of the symbol is the same as the duration of the chirp. Each symbol or chirp is composed of multiple samples, depending on the sampling rate and symbol duration. For instance, for sampling rate of $10^6$ samples per second, a symbol duration of $2$ ms corresponds to 2000 samples.

%% file: challenges.tex
\section{Challenges}\label{sec:challenges}
As mentioned before, \name\ aims to achieve the triple objectives of: sensitivity, real-time operation, and compatibility. However, each of these objectives is challenging on its own.

First, let us consider sensitivity. Sensitivity of LPWANs protocols is related to the bandwidth. Lower BW signals experience less noise and hence, can be received at lower signal strengths. Conversely, higher BW signals require higher signal strength at the receiver to be correctly decoded. Therefore, if \name\ configures its SDR to operate at a low BW, it meets the sensitivity requirements, but misses out on signals received at higher BWs. On the other hand, if \name\ sets its BW too high, it will miss out on signals coming from longer distances (and hence lower signal strength) at lower BWs.

Second, to ensure real-time operation, the SDR needs to identify the correct configuration of the packet using just a few symbols. However, the length of the symbol itself depends on the configuration used by the transmitter. A symbol sent using SF of 12 is 64 times longer than a symbol using SF of 6. If we capture the signal too long, we risk missing out on an entire packet for the highest data rate senders. On the other hand, if we capture it too short, we might not have enough information to identify the right configuration for the low data rate senders. 
\begin{figure*}[t]
    \centering	
    \begin{subfigure}[b]{0.3\textwidth}
    \includegraphics[width=\textwidth]{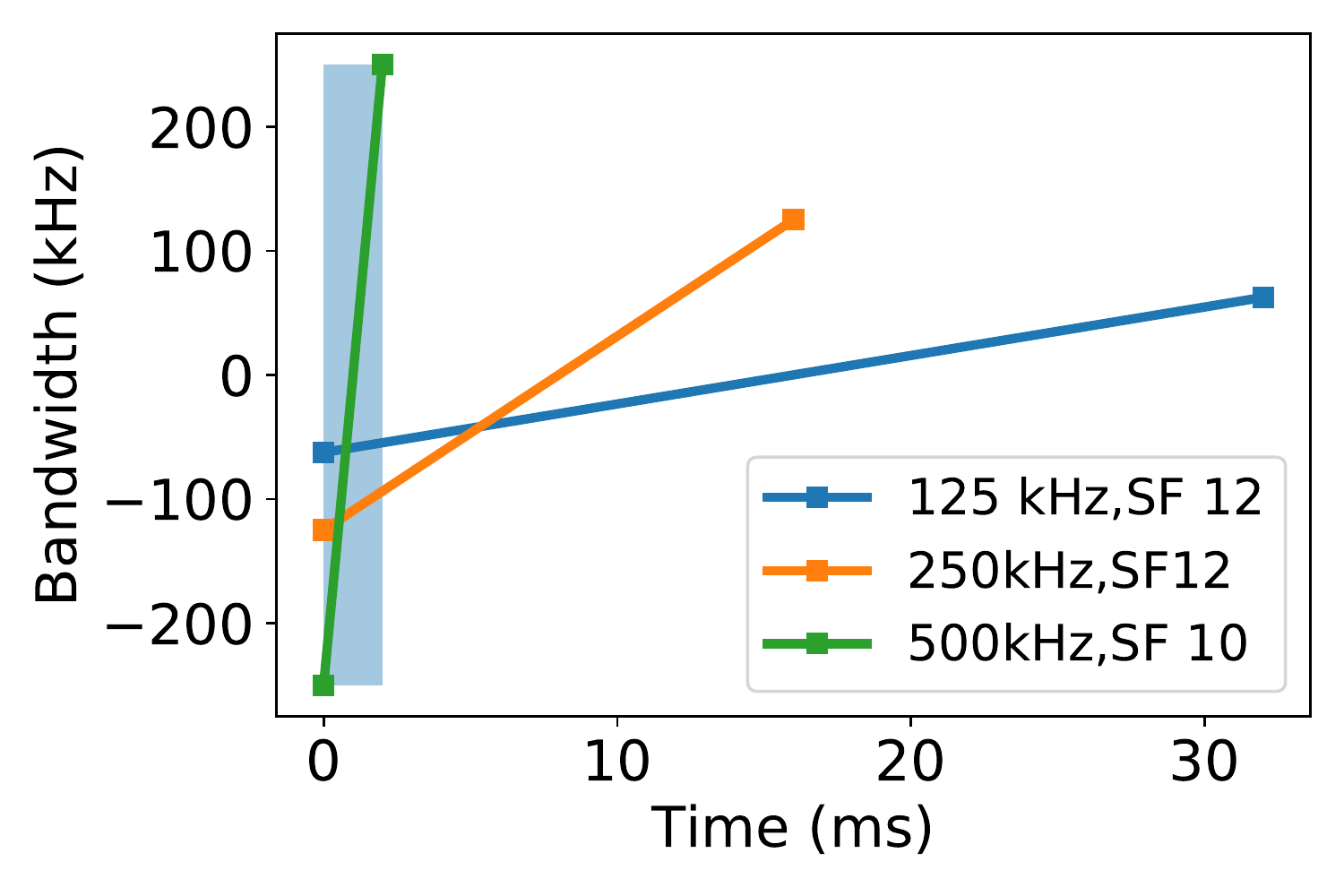}
    \caption{Sampling for high BW, low SF}
    \label{fig:sample_bw}
    \end{subfigure}
    \quad
    \begin{subfigure}[b]{0.3\textwidth}
    \includegraphics[width=\textwidth]{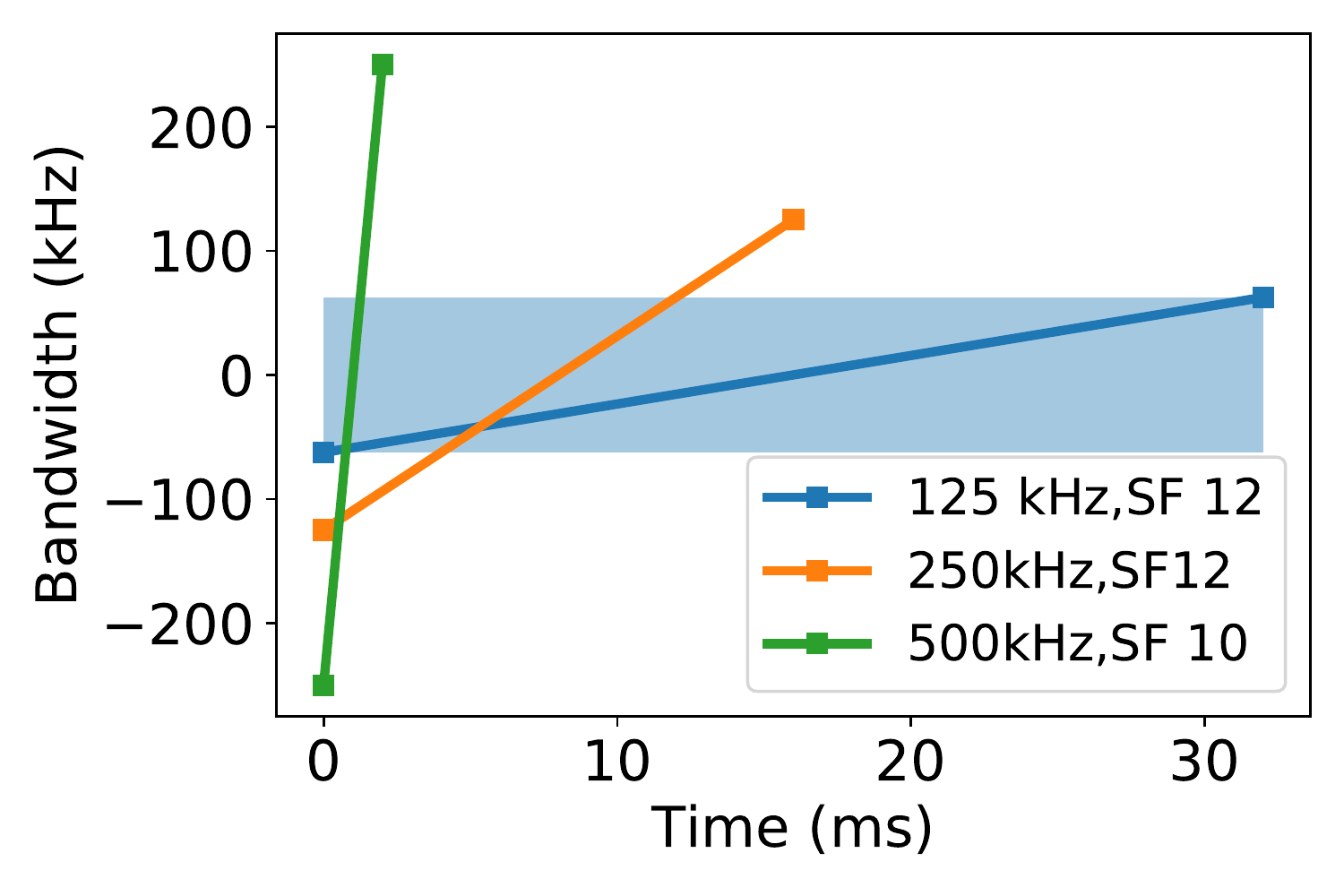}
    \caption{Sampling for low BW, high SF}
    \label{fig:sample_freq}
    \end{subfigure}
    \quad
    \begin{subfigure}[b]{0.3\textwidth}
    \includegraphics[width=\textwidth]{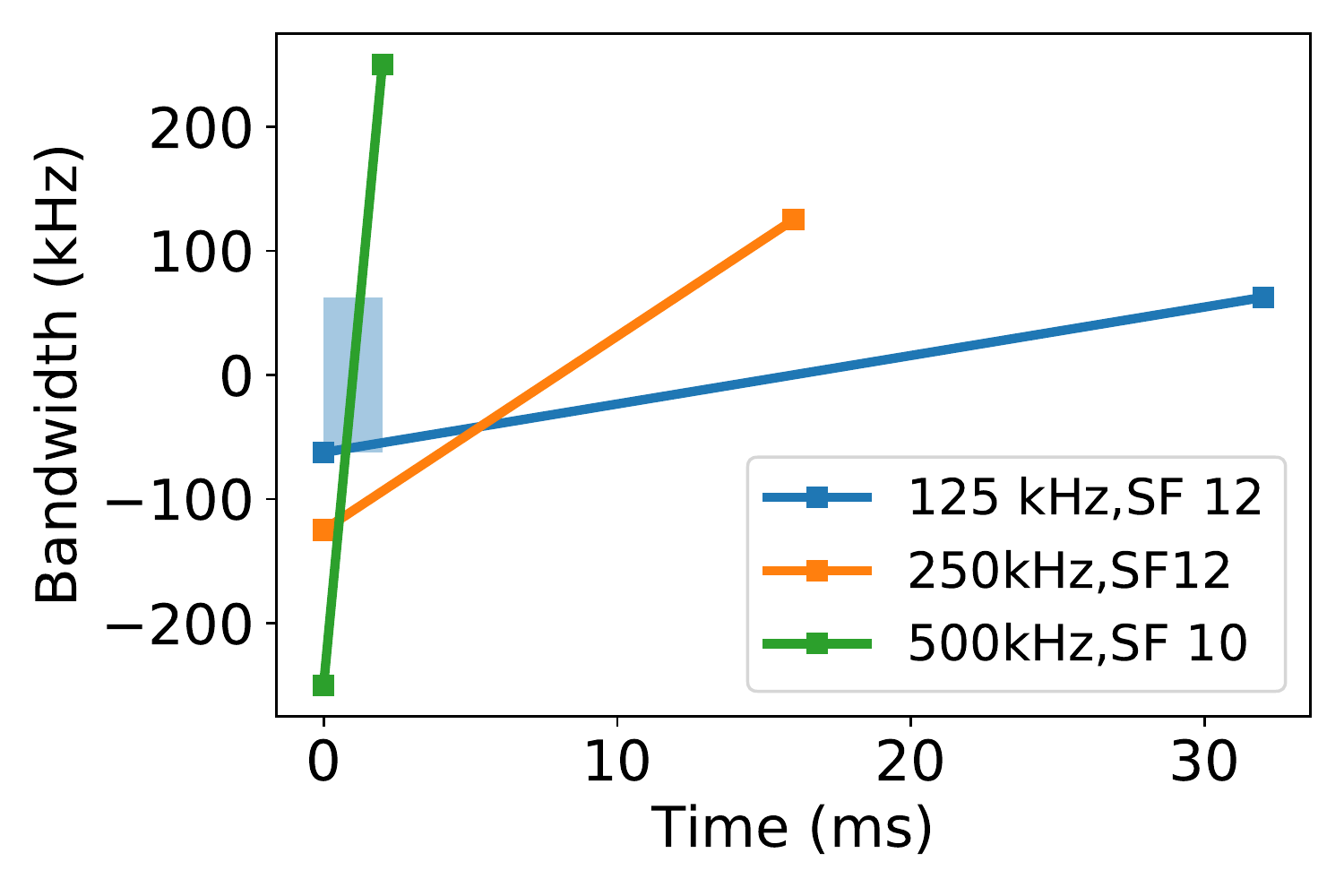}
    \caption{Sampling for low BW, low SF}
    \label{fig:sample_min}
    \end{subfigure}
    \vspace{-0.15in}
    \caption{\textbf{Sampling Challenge:} \textmd{The challenges with choosing the number of samples to use with respect to BW. Sampling (a) an entire symbol for the largest BW decreases sensitivity and  misses out on data for large SF settings. Sampling (b) at a low BW and entire symbol duration increases sensitivity but also introduces a large increase in latency. Lastly, sampling (c) at the minimum BW and symbol duration misses out on entire configurations.}}
    \vspace{-0.15in}
    \label{fig:sample_challenge}
\end{figure*}

Finally, to ensure backwards compatibility, the system needs to receive the \textit{entire packet} after the right configuration has been set on the gateway. However, that would mean we identify the configuration \textit{before} the signal even reaches the gateway -- a task that would seem impossible at first blush.

Fig.~\ref{fig:sample_challenge} demonstrates the challenge associated with configuring the receiving SDR itself. The figure shows chirps for three different configurations, that are relatively close to each other (it would be difficult to visualize more stark differences on the plots). As shown in Fig.~\ref{fig:sample_bw}, sampling one symbol of large BW and low SF configuration (highlighted in blue) has low sensitivity (due to high BW) and leaves us with a small temporal segment of the longer symbols. On the other hand, if we sample one symbol length for the low data rate configuration in Fig.~\ref{fig:sample_freq}, we maintain high sensitivity, but it introduces multiple symbol latency for the high data rate symbol. Finally, one might wonder, why we don't take the minimum of both frequency bandwidth and symbol duration across all possible configurations. This will ensure both sensitivity to low signal strengths and real-time operation. However, as shown in Fig.~\ref{fig:sample_min} such a configuration will end up missing out on some configurations all together. 

To resolve this conflict between sensitivity and real-time operation, \name\ takes a multi-stage approach (Fig.~\ref{fig:filter_bank}). It uses a set of band-pass filters in the digital domain to sample small chunks of BW for a short duration of time. It uses these small chunks of BW and frequency to decide if it has captured a signal long enough to make a decision about the configuration or if it needs to sample more. In no instance does \name\ use more than a duration of two symbols for any configuration to make this decision. 

Finally, in order to be compatible with existing hardware, the gateway needs to receive the entire packet \textit{after} it has been configured. Since the SDR is using at least some part of the preamble to identify the gateway, this goal seems impossible at first glance. One way to solve this problem is to buffer the time sample at the {\name} gateway, and then replay it at the gateway. However, this complicates the {\name} circuitry, and also drives up its cost. Instead, to solve this challenge we exploit the structure of the preamble in the LoRa protocol. Our key insight is that the preamble length of a packet can be dynamically configured. And it can be larger than the preamble length needed by a gateway to detect the packet. The rest of the symbols can be used by {\name} to determine the configuration parameters, and to set these parameters at the gateway radio. For instance, the gateway can be configured to expect a preamble of 8 symbols, but the client can be configured to use 10 symbols. These extra two symbols can be allocated towards \name\ to predict and reconfigure the LoRa gateway. The gateway can then use the remaining signal to decode the packet. Note that, since the number of upchirps is variable, the gateway still sees a full preamble with a sequence of upchirps followed by two downchirps and can successfully decode the packets.

%% file: method-3.tex
\section{\name}\label{sec:method}
{\name} is a new gateway design for LoRa that supports per-packet link configurations. With \name, clients can optimize their data rates without needing to inform the gateway of their updated configuration. In turn, this allows a single gateway to support 100s of endpoint devices without compromising performance. For example, a LoRa network deployment can include client devices that are dispersed across several miles from the gateway. Across this coverage area, the achievable throughput varies with respect to distance and diverse channel conditions. \name\ enables each of the clients to operate at their best configuration. %Today, which otherwise would have to compromise performance in order to support all devices in the vast coverage area. 

\begin{figure}
    \includegraphics[width=\columnwidth]{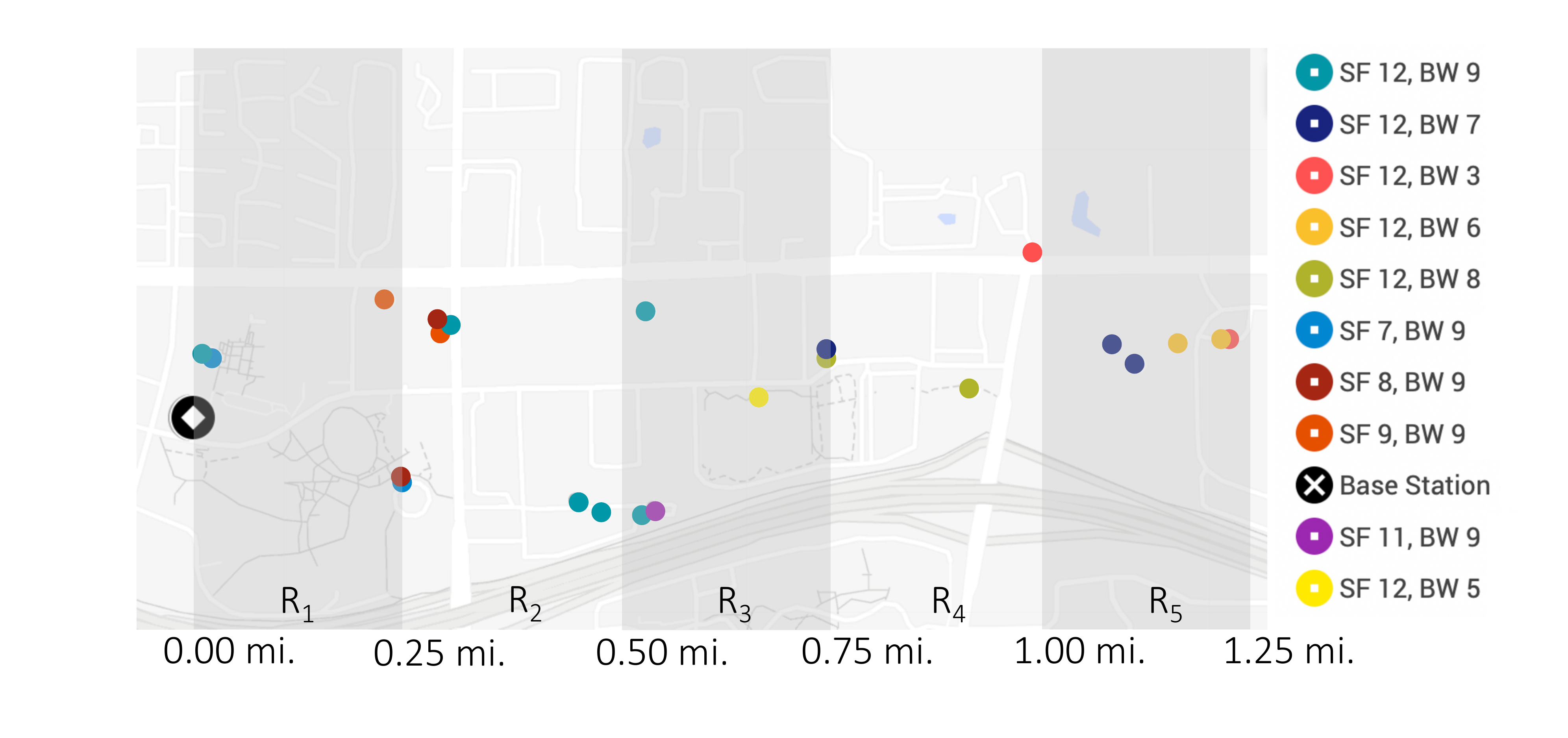}
    \caption{Coverage Map for LoRa. \textmd{We perform measurements using an off the shelf LoRa chipset on an industrial complex. This figure denotes the best bandwidth, spreading factor combination that can be supported on each location.}}
    \label{fig:map}
\end{figure}

To better understand the performance of LoRa, we conduct a range test and determine the maximum achievable data rate with respect to distance from the gateway. Fig.~\ref{fig:map} shows a coverage map for the best configuration settings that can be reliably supported. The gateway was placed at a fixed location and the client location was varied across a campus. The client continuously transmitted LoRa packets with a transmit power of 20dB and at each location we varied encoding parameters to identify the best configuration. The figure shows the maximum supported data rate across all locations as well as the corresponding BW and SF. The key takeaway is that there is much variation across encoding parameters, justifying the need to support a dynamic network. 

\name\ achieves this by taking a neural networking (NN) approach to predict the bandwidth and spreading factor used by any given client for data transmission. In turn, the LoRa gateway is reconfigured accordingly to decode incoming packets. The architecture of the \name\ gateway is shown in Fig.~\ref{fig:arch}, which has three components: a packet detection module to detect incoming LoRa packet transmissions, a NN processing unit to classify the encoding configuration, and lastly a parameter configuration module that communicates with the LoRaWAN gateway to update encoding parameters. 

Bringing this design to fruition requires addressing key challenges highlighted in Section ~\ref{sec:challenges}. First, \name\ needs to determine the configuration parameters of a received packet in near real-time. Second, it needs to be backwards compatible with existing LoRa solutions, and finally, \name\ must achieve high prediction accuracy across all configuration settings. Below, we detail how \name\ addresses these challenges.

\subsection{Real-time Prediction}
Recall, we have set an ambitious goal for \name: to enable per-packet reconfiguration at the gateway. To enable such reconfiguration, the \name\ gateway must determine the configuration without losing any packet data content.

Our insight is that we can vary the number of upchirps in the LoRa preamble without altering the LoRa protocol and hardware. Specifically, if we configure the client to transmit $N$ upchirps and the gateway to expect $N-k$ upchirps in the preamble, then we can use the $k$ extra up-chirps to identify and set the correct configuration parameters, without altering the gateway or client hardware.

To validate this approach, we configured two Semtech SX1276 LoRa chips as a gateway and client. We use the API of popular LoRa chipsets (Semtech SX1262/1276) to configure the preamble of a LoRa packet from 6-65535 symbols, where the the minimum of 6 symbols is needed for packet detection. We set the preamble at the gateway to 8 symbols, while varying the client's preamble length. The client transmitted packets over the air using different preamble lengths which are received by the gateway. Our experiments demonstrate that an extra 5 symbols can be added to the preamble of the client device, while still maintaining reliable reception at the gateway. Given this observation, we have a budget of $k=5$ symbols for detecting and setting the parameters.

Finally, recall that the length of a symbol in time domain is a function of the BW and SF parameters themselves. Therefore, the time budget for decoding is the lowest time across all the configuration parameters possible.

%\name\ requires up to 2 symbols depending on the encoding parameters used for the input data. The variation is due to the fact that for any neural network the input shape of the data must be consistent. Depending on the number of data samples passed into the network for classification, the number of symbols used for any given input will vary as well because the duration of the symbol is a function of BW and SF. 

%\begin{figure}
%  \centering
%    \begin{subfigure}[b]{0.4\textwidth}   \includegraphics[width=1\textwidth]{figures/lora_BW1_features.pdf}
%    \caption{\footnotesize{\textmd{FFT for BW=10.4kHz and SF ranging from 10-12}}}
%    \label{fig:bw1_feat}
%    \end{subfigure}
%    \hspace{5pt}
%    \\
%    \begin{subfigure}[b]{0.4\textwidth}       \includegraphics[width=1\textwidth]{figures/lora_BW2_features.pdf}
 %   \caption{\footnotesize{\textmd{FFT for BW=15.6kHz and SF ranging from 10-12}}}
 %   \label{fig:bw2_feat}
 %   \end{subfigure}
 %   \hspace{5pt}
 %   \\
 %   \begin{subfigure}{0.4\textwidth}       \includegraphics[width=1\textwidth]{figures/lora_BW7_features.pdf}
 %   \caption{\footnotesize{\textmd{FFT for BW=125kHz and SF ranging from 10-12}}}
 %   \label{fig:bw7_feat}
  %  \end{subfigure}
  %  \vspace{-0.15in}
  %  \caption{Data Features Extracted from LoRa symbols. \textmd{The real and imaginary components as well as the FFT for LoRa symbols are used as features for \name\ to classify encoding parameter configurations.}}
  %  \vspace{-0.15in}
  %  \label{fig:features}
%\end{figure}

\subsection{Inferring SF and BW}
We can mathematically model the LoRa sample generation process as a function, $G(\theta,n)$, where $\theta={BW, SF, R, SNR}$ is the set of parameters bandwidth, spreading factor, sampling rate, and signal-to-noise ratio respectively, and $n$ is the number of samples that we intend to generate. For generating up-chirps around center frequency $f$, $G(\theta,n)$ starts at frequency $f-\frac{BW}{2}$ and ramps up the frequency to $f+\frac{BW}{2}$ over samples proportional to $2^{SF}$. LoRa devices implement this process in hardware. Our goal is to reverse engineer this process, to go from a sequence of samples to underlying parameters $BW$ and $SF$ used to generate the sequence.

At first blush, this problem seems trivial: the starting frequency of the chirp reveals the BW and the number of samples used for a single chirp reveal the SF. However, as mentioned before, \name's goal is to predict the parameters \textit{before} the data transmission begins. Since the actual chirp length can vary by several orders of magnitude depending on the parameters, it is impossible to sample the complete chirp before making this decision. This is because waiting to sample a complete chirp for the slowest modulation can consume over a thousand symbols for the fastest modulation completing missing packets at the high data rate. Therefore, we need to make decisions based on fractional chirps for most of the modulations. In fact, for some of the modulations, we must make this decision based on smaller than one-thousandth of the chirp.

This problem is further complicated by two other aspects of signal transmission: (a) frequency-selective multipath fading experienced by a chirp makes frequency and timing estimation challenging, (b) LoRa devices operate at low SNRs. Both of these factors induce additional noise in the measurement process. As we show in Sec.~\ref{sec:results}, even a strong baseline such as auto-correlation with the transmitted signal achieves poor accuracy on this task due to the enhanced noise and the need to make an estimate from a small number of samples.

We note that this problem is essentially a pattern classification problem that has been solved in other domains (like computer vision) successfully using Convolutional Neural Networks (CNN). In fact, past work like \cite{oshea} has used CNN's to classify modulation schemes like BPSK, QPSK, QAM, etc. using complex time-domain signals. Therefore, we opt for a CNN-based classification network design for \name. However, unlike past work, we must deal with partial symbols and varying bandwidths, which we discuss below.

%Instead, \name\ takes into account the described characteristics of LoRa chirps to train a convolutional neural network (CNN) to classify the many different combinations of SF and BW. Specifically, we use three features extracted from the symbols of the LoRa preamble to perform classification. The first two features are the real and imaginary components of the signal and the last is the Fast Fourier Transform (FFT). Using data from both the time and frequency domain of the signal is crucial in achieving high prediction accuracy. For instance, if we were only to rely on the FFT of each signal we would find it almost impossible to distinguish between the very low BW settings. As shown in Fig.~\ref{fig:features}, when evaluating the FFT for different BW and SF settings, the lower kHz range begins to look quite similar. Supplementing this with features from the time domain helps catch the variation in the frequency of oscillation of a preamble symbol, while the FFT helps provides insights on bandwidth variation.

\begin{figure}
    \centering
    \includegraphics[width=0.95\columnwidth]{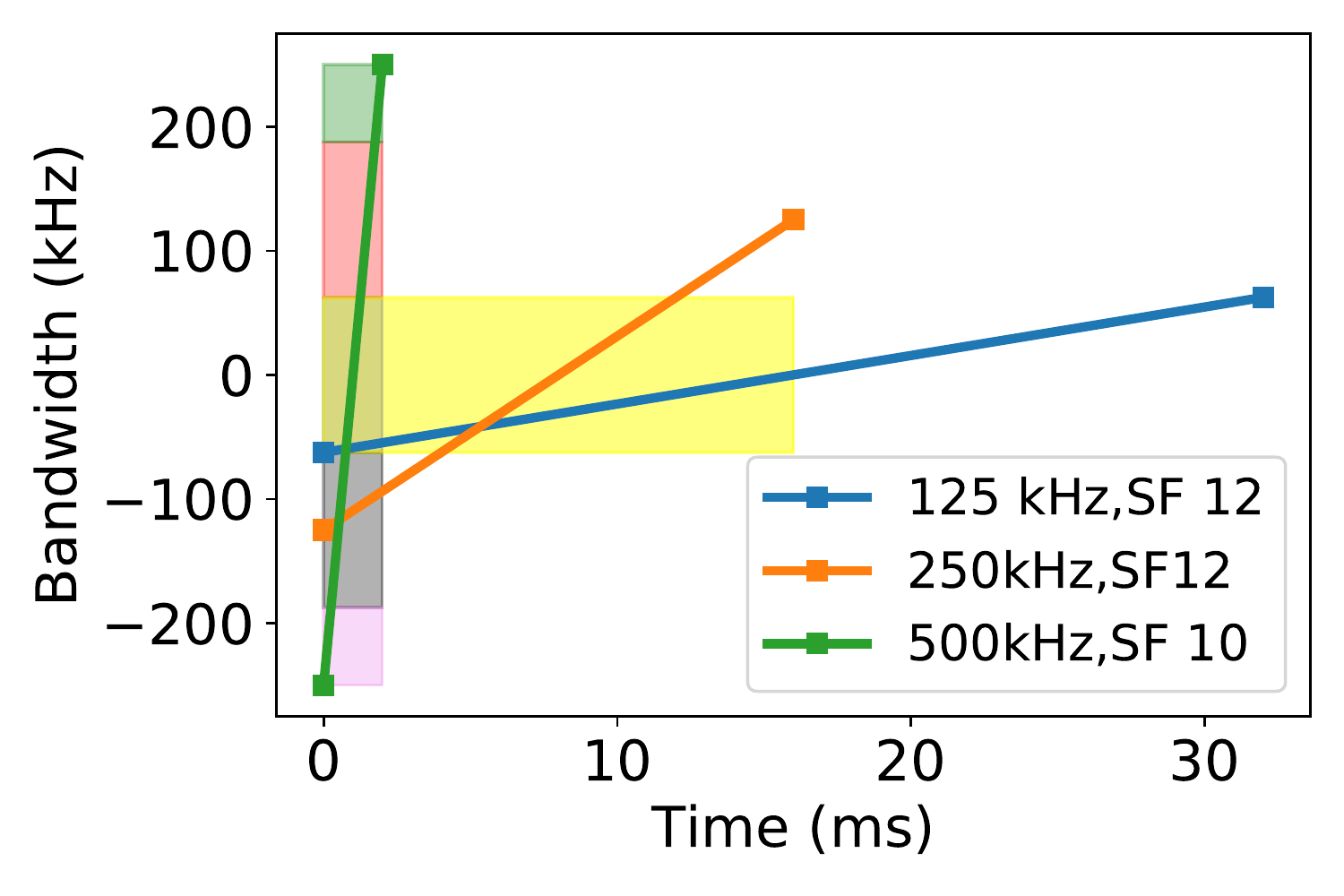}
    \vspace{-0.1in}
    \caption{\textbf{Multi-stage Filter Bank:} \textmd{By using a set of parallel narrowband filters, we retain sensitivity while sampling a larger band. Using a small set of samples in the beginning, \name\ decides if it needs to sample more or if it can make a decision on configuration. In no cases, it uses more than two symbol durations.}}
    \label{fig:filter_bank}
    \vspace{-0.1in}
\end{figure}

\subsection{Multi-Stage Filter Bank}
Recall, the chirp-length for lower data rates is several orders of magnitude smaller than the highest data rate. Furthermore, the highest bandwidth modulation operates at a high SNR which is not suitable for the lowest bandwidth modulation. This structure poses a fundamental problem: how should a \name\ receiver operate \textit{before} it has identified the right configuration? Does it operate at a high bandwidth and risk losing the lowest bandwidth transmissions due to SNR or does it operate a low bandwidth signal and risk losing information about the highest bandwidth transmissions? Similarly, should we sample for a long time period to get a large fraction of the low data rate transmission, but add impractical latency to the high data rate transmissions or should we sample a tiny fraction of the low data rate symbols and risk losing them due to low SNR?

Clearly, none of these choices are desirable. Therefore, we propose a multi-stage filter bank approach. For bandwidth, we sample a wideband signal and add multiple digital narrowband band pass filters on top to retain sensitivity to low SNRs. This allows us to capture the entirety of the wideband signal while maintaining low SNR functioning for low bandwidth symbols. The number of narrowband filters and the size of each filter is decided by the specification of the protocol. For the LoRa protocol, we choose 10 filters of 20 kHz each.

\begin{figure}
\vspace{0pt}
\centering
\includegraphics[width = 0.4\textwidth]{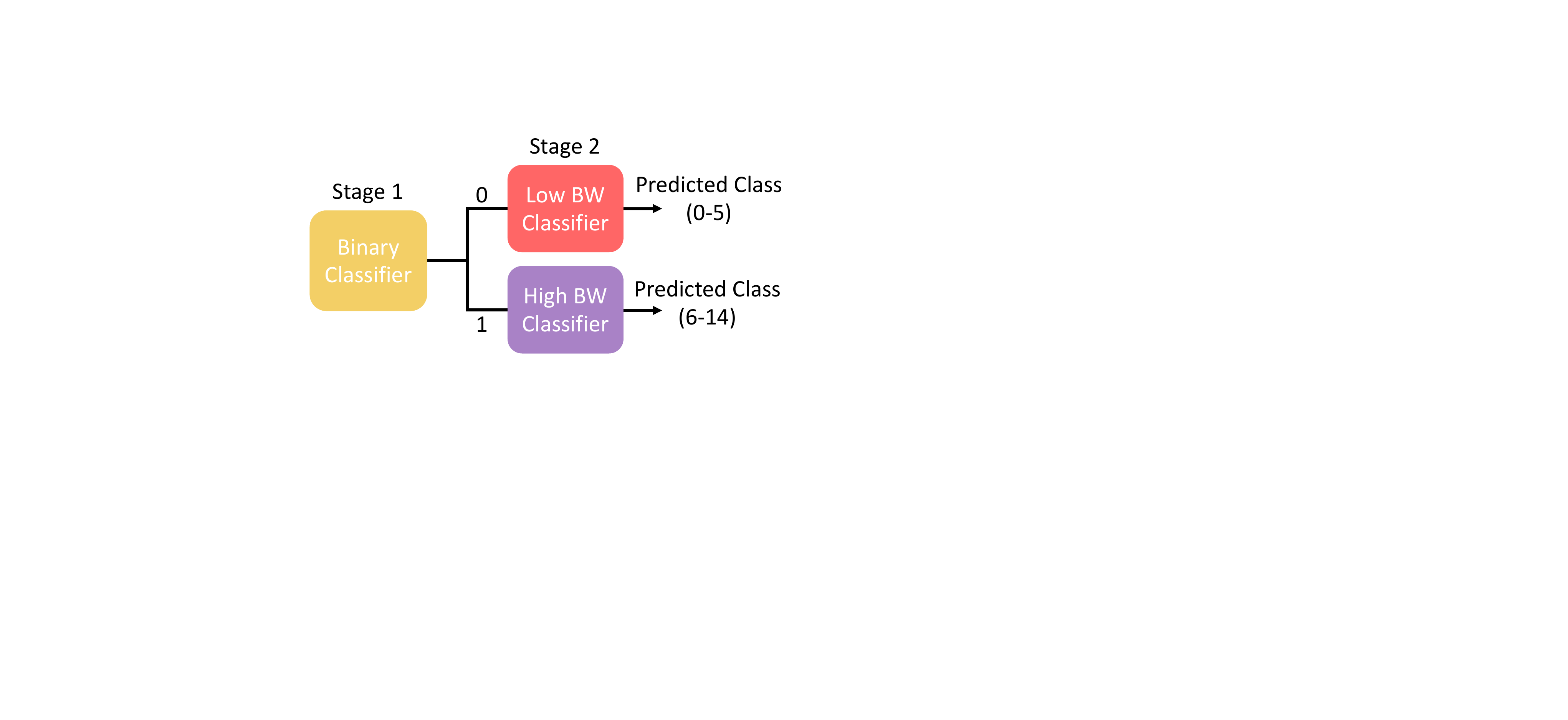}
\vspace{-0.0in}
\caption{Hierarchical model.\textmd{\name\ uses a two stage CNN to perform classification. First, a binary classifier to distinguish between lower and higher BW and followed by classifiers that identify the BW-SF combination.}}
\label{fig:hier}
\vspace{-0.0in}
%\vspace{3pt}
\end{figure}

To solve the time sampling problem, \name\ uses a multi-stage approach. Specifically, we manually segment the different modulations into two categories: high data rate and low data rate. We begin by sampling a duration of just 2 chirps at the highest data rate. \name\ uses this small sample set (4 ms) to decide either (a) this is a high data rate chirp and must be classified based on the current samples, or (b) this is a low data rate chirp and we must sample a longer fraction of the chirp (65 ms) and collect more data before making a decision. This yields to a multi-stage approach shown in Fig.~\ref{fig:filter_bank} and Fig.~\ref{fig:hier}.

Note that, the segmentation of modulations into high data rate (BW>20 kHz) and low data rate (BW<20 kHz) is based on the current LoRa protocol parameters implemented in frequently used hardware. As the number of modulations increase, this segmentation can be hierarchical and follow a tree structure. We leave this investigation to future work.

 %\name\ uses a multi-stage sampling method to optimize sensitivity, latency, and classification accuracy. We can refer back to Fig.~\ref{fig:filter_bank} which illustrates the adaptive approach. We use a set of digital band-pass filters to create subsets of bandwidth for a short time duration. These subsets are used to determine if the captured signal is long enough to provide accurate insights on radio configurations or if we should continue sampling. In particular, \name\ uses a total of 12808 samples (65ms) for the first six class representing the two low bandwidths and 800 samples (4ms) for the last nine classes representing the higher bandwidth radio configurations. Intuitively, using a larger set of samples for the lower bandwidth settings makes sense since the symbol duration increases as BW decreases. 

\subsection{Neural Network Architecture}
\begin{figure*}
\vspace{0pt}
\centering
\includegraphics[width = 1\textwidth]{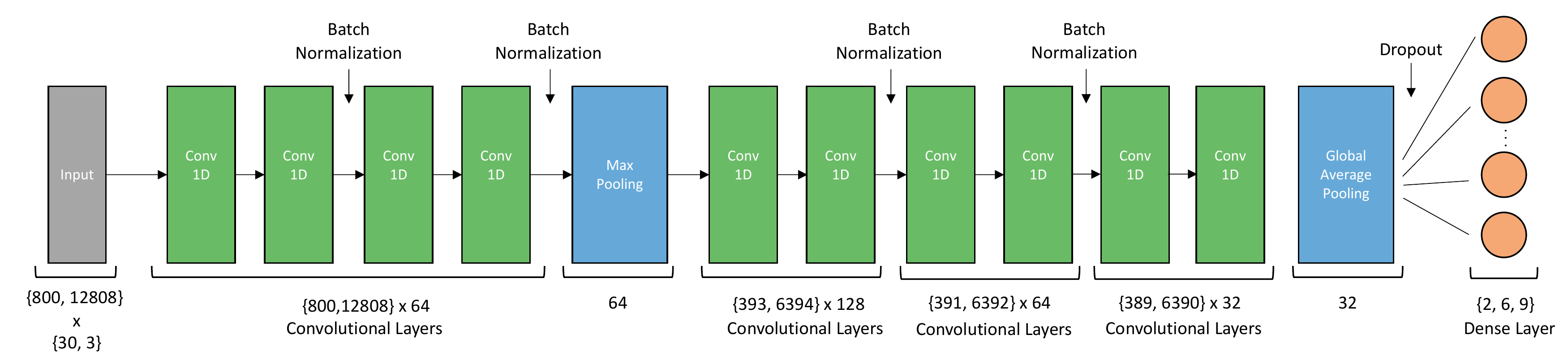}
\vspace{-0.2in}
\caption{\name\ Neural Network. \textmd{\name\ uses a CNN to classify encoding configurations of LoRa packet transmissions. It uses a combination of convolutional, max pooling, batch normalization, \& dense layers.}}

\label{fig:network}
\vspace{-0.15in}
%\vspace{3pt}
\end{figure*}

We use an identical architecture of the neural network for the different stages of the multi-stage classification process. The only difference is the number of output classes and the number of input samples. We decide these parameters depending on the stage of the network. Fig.~\ref{fig:network} illustrates the NN architecture. In principle, the goal of this architecture is to model the function $G^{-1}$ that takes in a set of samples and returns the parameters used to generate these samples.

%\name\ uses a hierarchical NN architecture that includes two stages. As shown in Fig.~\ref{fig:hier}, we first use a binary classifier to distinguish between the lower and higher bandwidths. Depending on the prediction, this will follow by either a six or nine class classifier to predict the BW and SF radio configurations. Fig.~\ref{fig:network} illustrates the NN architecture used by \name\ at each stage, where the key difference is the number of classes, features, and samples inputted per classifier. 

\vspace{6pt}\noindent\textbf{Features:} The \name\ architecture uses the output of the multi-stage filter bank as the input to the NN. The input is shaped $N_{t} \times N_f \times 3$, where $N_t$ is the number of samples that is dependent on the stage of the network. For the bandwidth classifier (binary) and the high bandwidth classifier, we set $N_t=800$, corresponding to 4 ms of data (two times the chirp length for the fastest data rate). For the low bandwidth classifier, $N_t = 12808$, corresponding to 64 ms of data (fastest chirp rate among the low bandwidth configurations). $N_f$ is set to the number of filters used. For high bandwidth and binary classifier, $N_f=10$ because we need to access the complete wideband signal. For low bandwidth, $N_f=1$ because we just need to access one narrowband signal, the other parts are just noise or interference. Finally, we perform the FFT on the signal along the $N_t$ axis and use the real, imaginary, and absolute value of the signal as the input to the NN.

%The low bandwidth classifier relies on the three features described previously. The binary and nine class classifier uses 30 features to predict radio configurations. The features include the real and imaginary components and FFT; however, the samples are split into ten 20kHz chunks. The variation in the number of features and samples per classifier is chosen based on the type of signals that need to be classified, as previously discussed in Sec.~\ref{sec:challenges}. For instance, we use 16x more samples for the low bandwidth classifier because the symbol duration can be 10s of milliseconds and more samples are needed to have a meaningful feature. On the other hand, the binary classifier uses only 800 samples even for the low bandwidths, but since it does not need to distinguish between individual bandwidths this is sufficient. 

%We use a 20kHz bandpass filter to represent a 20kHz bandwidth setting used at the receiver that gives us better sensitivity. A challenge with using such a low receive bandwidth is that meaningful data is not captured for the higher bandwidths. Instead, we filter in the digital domain where the signals received with a bandwidth of 200kHz are spilt into ten 20kHz chunks. 

\vspace{6pt}\noindent\textbf{Network Design:} The NN for each classifier starts off with four convolutional layers, each with a filter size of 128. The layers convolve the input and are activated by a Rectified Linear Unit (ReLu) function. The ReLu activation function outputs the max of zero and the input data and provides an output in the form of a feature map. Next is a max-pooling layer that is used to reduce the size of the generated feature map and retain the most meaningful information. In our NN we use a max-pooling size of two. This is followed by six more convolutional layers each with a filter size ranging from 128-32. These layers also use the ReLu activation function. A global average pooling layer is added after this, which calculates the average outputs of each feature map from the previous convolutional layer. We apply a final dense layer, whose size is equal to total number of possible classifications. The dense layer uses a sigmoid activation function that provides the output probabilities across all classes between values of 0 and 1. To retrieve the the predicted class we simply take maximum probability of the final output layer. 

Within the NN, we also add three batch normalization layers and a dropout layer. The batch normalization layers normalize the output of the previous layer by subtracting the batch mean and dividing by the batch standard deviation, where a batch is a portion of data passed into the model for training~\cite{batch_norm}. Batch normalization improves the stability of the NN and helps in reducing the number of epochs needed to train the NN. Lastly, for regularization, we use a dropout of 0.5 to reduce overfitting.

\vspace{6pt}\noindent\textbf{Loss Function:} We evaluate how well the NN models our dataset by using a categorical cross entropy loss function,
\begin{equation}\vspace{-0.1in}
    -\sum_{c=1}^{N} t_{i,c}log(p_{i,c})
\end{equation}
where N is the number of classes, p is the predicted probability of the current sample, and t is the binary indicator of whether the class, c, is correct. The loss function evaluates performance of a classification model for output probabilities between 0 and 1. In other words, the cross entropy will increase if the model predictions stray from the actual value and in turn provides a measure for error. To have accurate predictions we also need to minimize the error and this is typically done by using an optimization function. At a high level, optimization functions calculate the partial derivative of the loss function with respect to weights used in the model. These weights are modified until a minima is reached for the loss function. \name's network architecture uses an Adam optimizer to perform this task~\cite{adam}.

%% file: implementation.tex
\section{Implementation}
We present details about \name's implementation below.

\subsection{Hardware} \label{sec:hardware}
We design a hardware prototype of the \name\ gateway using the universal software radio platform (USRP). The gateway operates at 915 MHz, the frequency used by LoRa deployments in the United States. The USRP is co-located with a LoRa receiver which needs to be set to the correct configuration for it to decode packets.
\begin{figure}[t!]
\vspace{0pt}
\centering
\includegraphics[width = 0.4\textwidth]{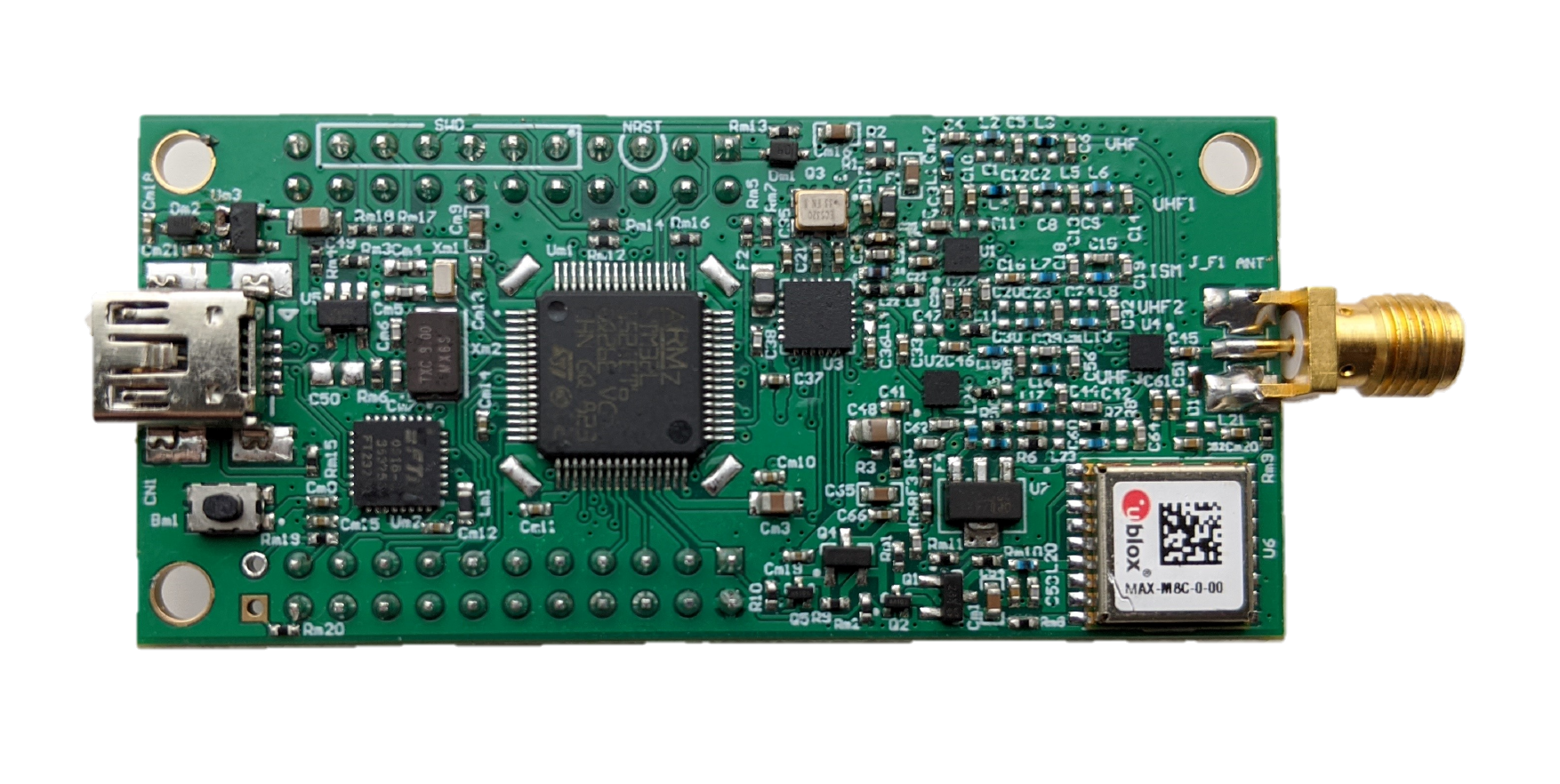}
\vspace{-0.2in}
\caption{Radio with integrated SX1276 LoRa chip. \textmd{A custom PCB designed to interface with the SX1276 to communicate using the LoRa protocol.}}
\vspace{-0.15in}
\label{fig:radio}
\end{figure}

The clients are designed using the 1276 Semtech chipset \cite{sx1276}. This chipset allows us to choose SF ranging from 7 to 12 and the bandwidth ranging from 7.8 to 500 kHz. We choose BW of 10.4, 15.6, 125, 250, and 500 kHz for our experiments to cover the extreme ends of the spectrum. Note that, we also cover the smallest difference between bandwidths by selecting two of the lowest bandwidths possible. Lastly, we use spreading factors of 10 to 12 for our experiments, as they are most commonly used spreading factors for very large-scale deployments.

The client chip is embedded in a PCB that sets the SF, BW, and allows us to write data bits to be transmitted, as shown in Fig.~\ref{fig:radio}. The chip is controlled using the ARM STM32L151 \cite{stm} micro-controller. We write custom firmware for this micro-controller. Note that this practice of embedding the off-the-shelf LoRa chip into PCB deployments is common for LoRa deployments. \name\ can operate with any such implementation on the client side without any modifications.

\subsection{Software}
We control the \name\ gateway using GNU Radio software \cite{gnu}. The software collects samples with a center frequency of 915 MHz and a sampling rate of 200ksps. This is the minimum sampling rate that can be achieved by the USRP and results in a 200kHz bandwidth at the receiver. Each packet recording is passed through a bandpass filter to further reduce the receiver bandwidth to 20kHz. The additional filtering is performed to increase the sensitivity of the receiver. The samples are then segmented into individual symbols using a packet detection algorithm that relies on a combination of power thresholding on a sliding window and auto-correlation.

We implement the CNN using the Tensorflow 2.0 framework \cite{tf} in Python. It runs on a Microsoft Surface 2 with 16 GB RAM and NVIDIA GeForce GTX 1050 GPU with 2 GB memory. The CNN is trained using the Adam Optimizer with default parameters aside from the learning rate set to 0.0001. 20 percent of the training set is set aside as a validation set. We train the model for \textbf{20} epochs for all of our experiments and choose the best model based on validation set performance. For each experiment, we run three different training-test splits, unless stated otherwise. The inference takes about 3 ms per prediction in our current implementation (less than 2 chirp lengths at the fastest data rate). This can be improved further by integration into the FPGA of the SDR, which we leave to future work.

%% file: results.tex
\begin{figure*}[t]
    \centering	
    
    \begin{subfigure}[b]{0.3\textwidth}
    \includegraphics[width=\textwidth]{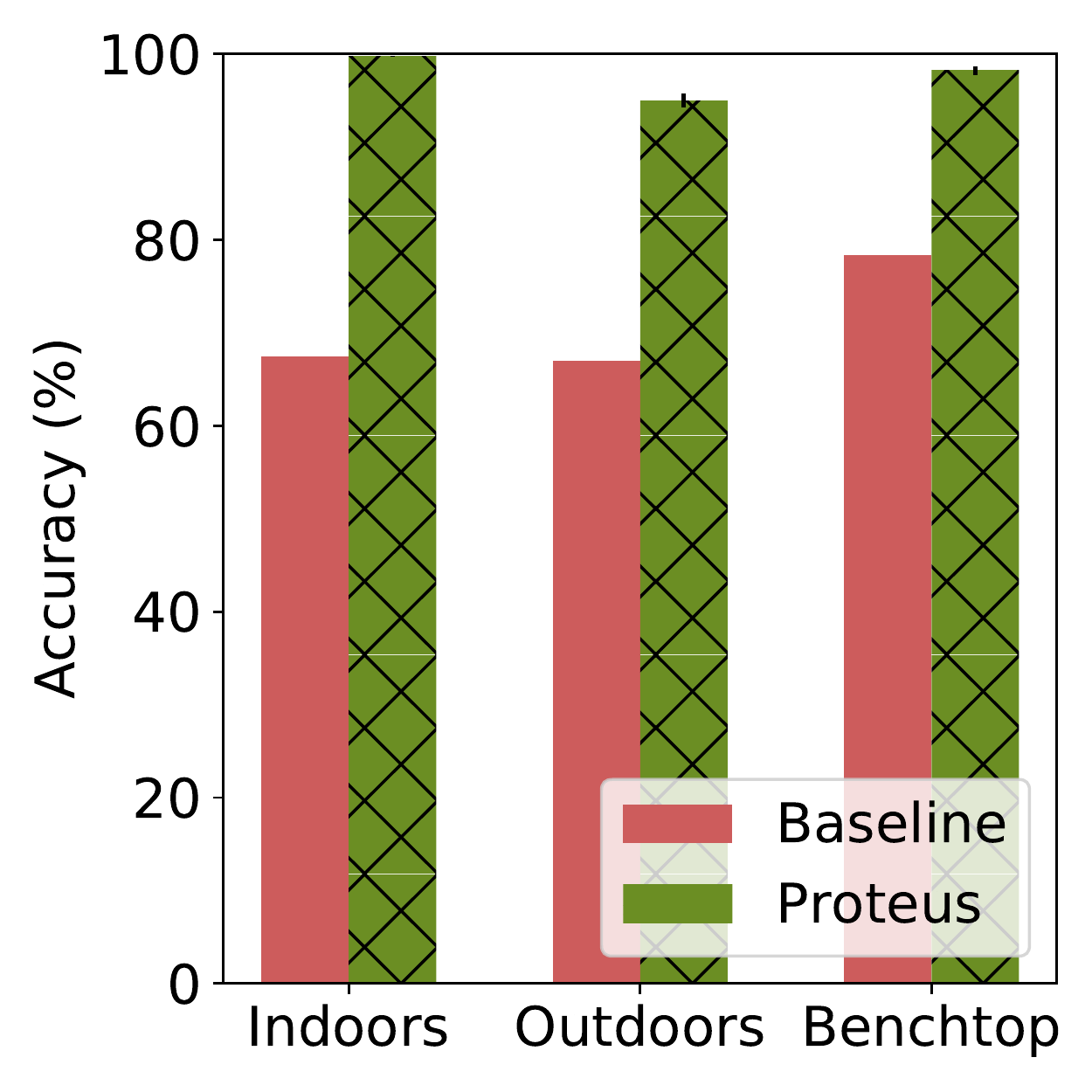}
    \caption{Accuracy Comparison}
    \label{fig:accuracy}
    \end{subfigure}
    \begin{subfigure}[b]{0.3\textwidth}
    \includegraphics[width=\textwidth]{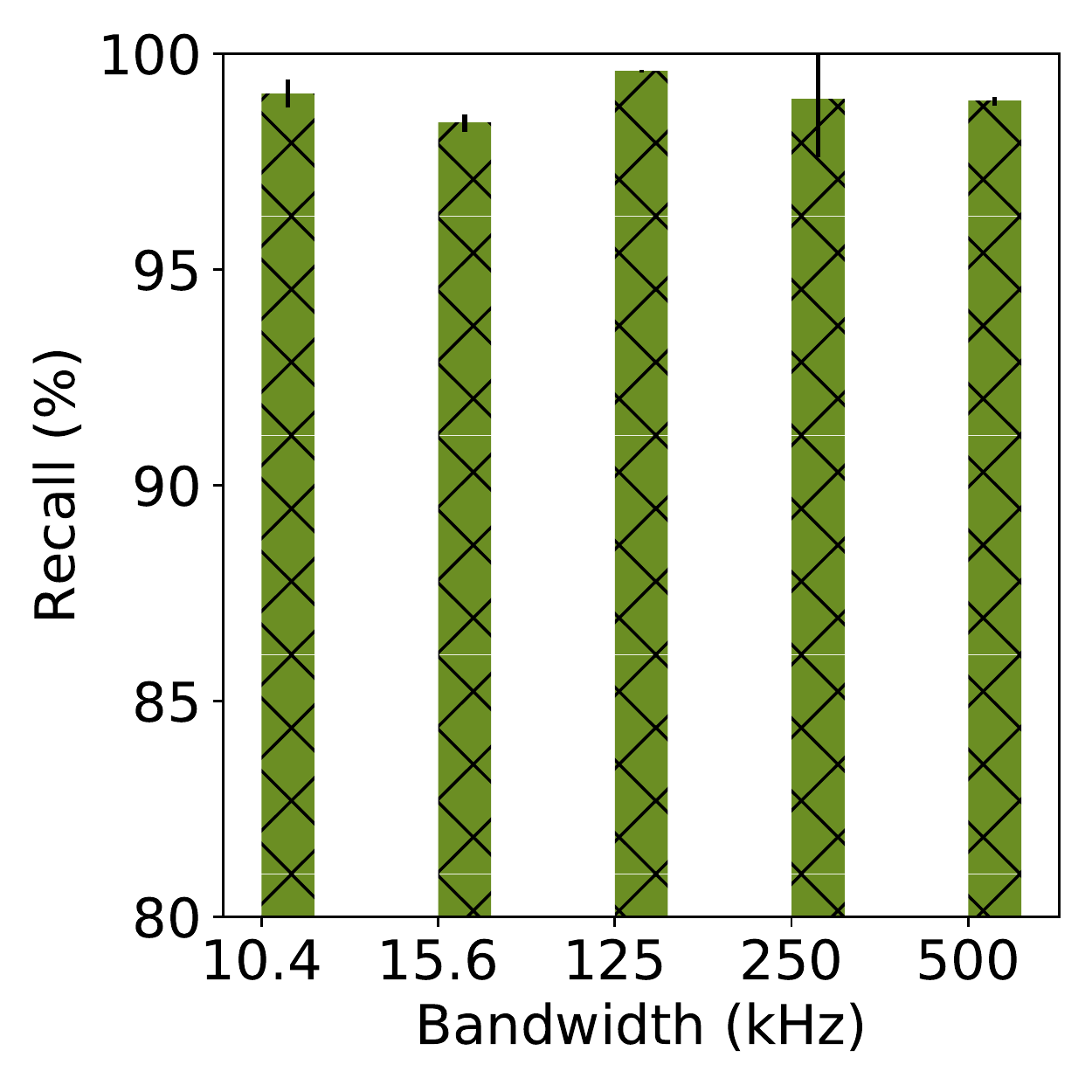}
    \caption{Variation with BW}
    \label{fig:accuracy_bw}
    \end{subfigure}
    \quad
    \begin{subfigure}[b]{0.3\textwidth}
    \includegraphics[width=\textwidth]{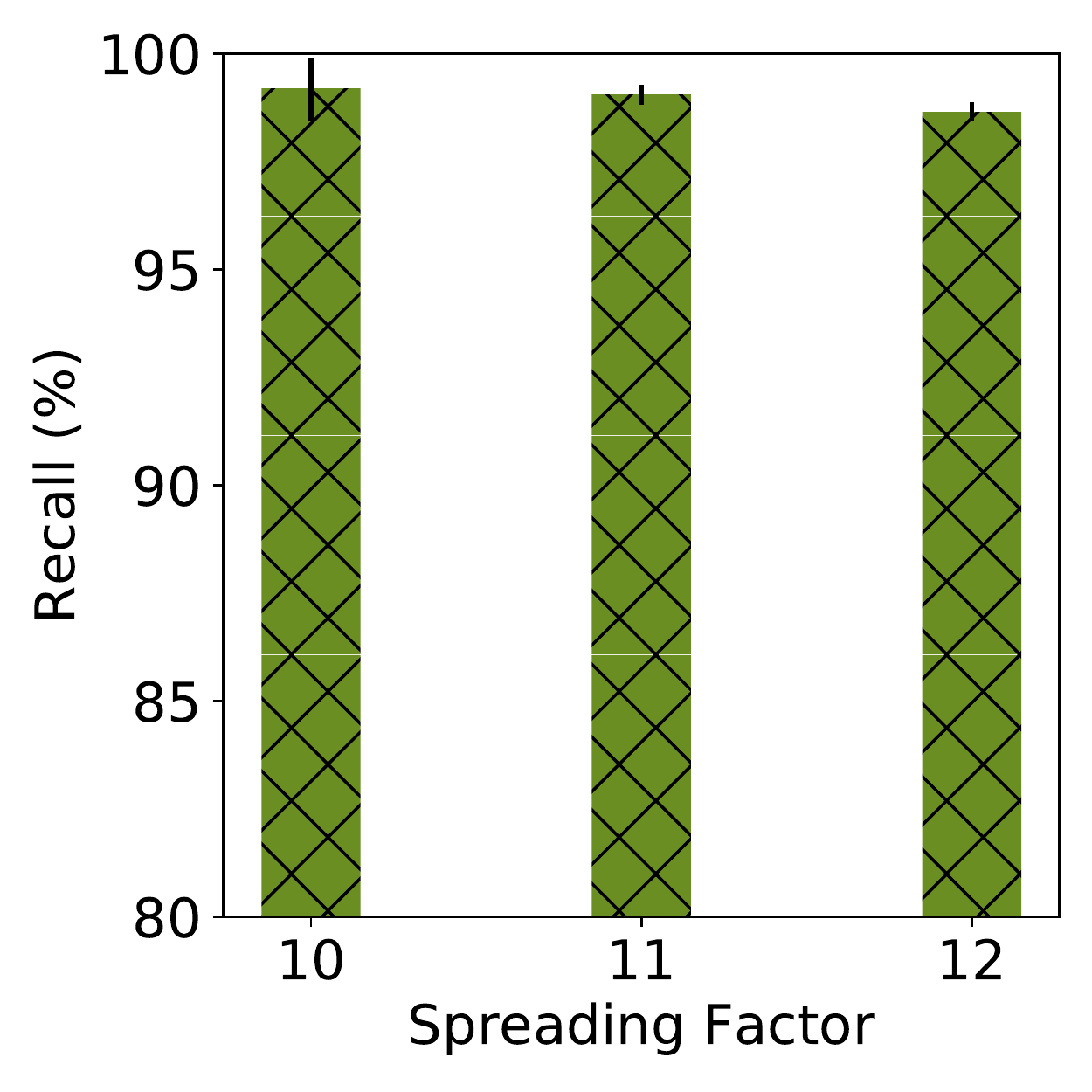}
    \caption{Variation with SF}
    \label{fig:accuracy_sf}
    \end{subfigure}
    \begin{subfigure}[b]{0.3\textwidth}
    \includegraphics[width=\textwidth]{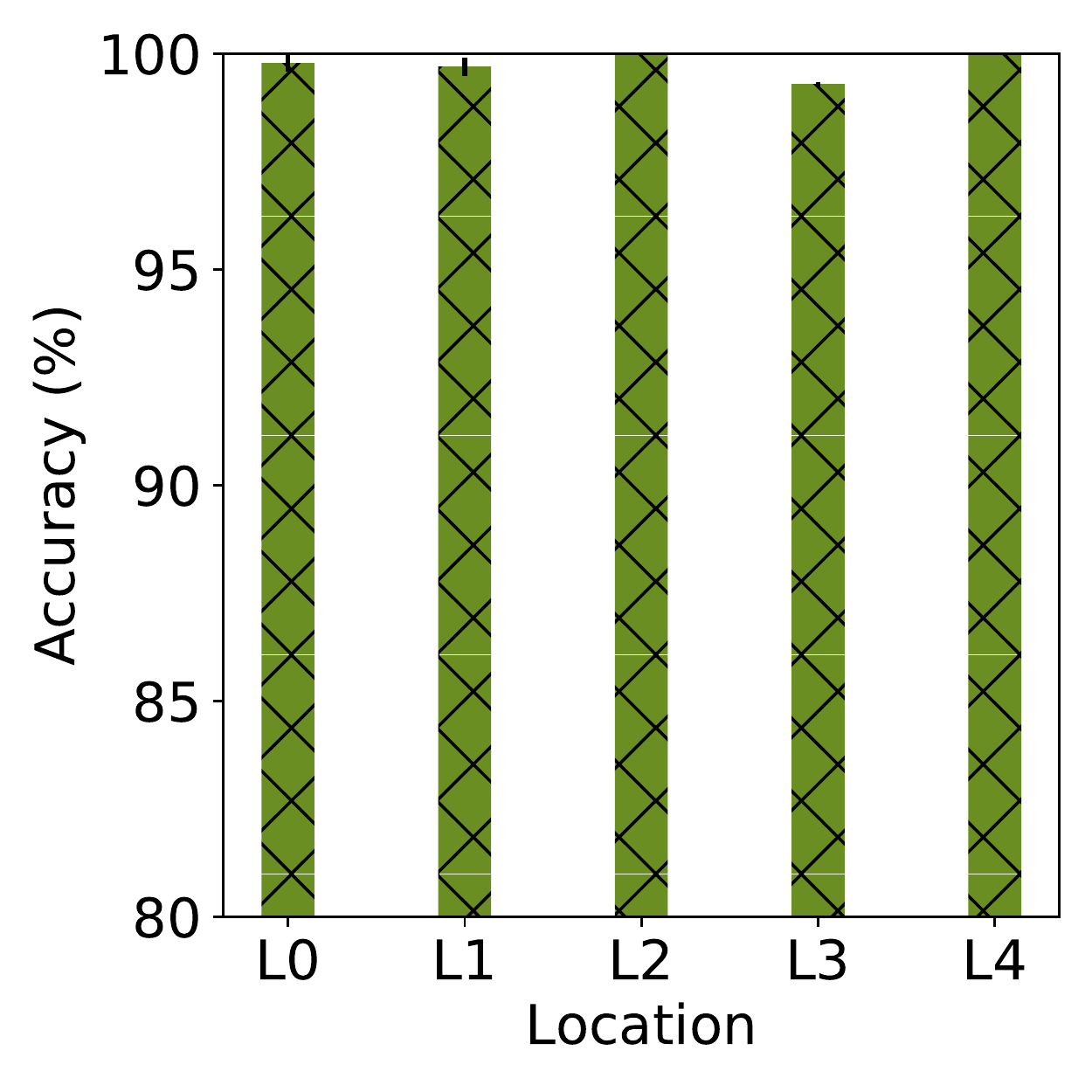}
    \caption{Variation with Location}
    \label{fig:accuracy_location}
    \end{subfigure}
    %\quad
    \begin{subfigure}[b]{0.3\textwidth}
    \includegraphics[width=\textwidth]{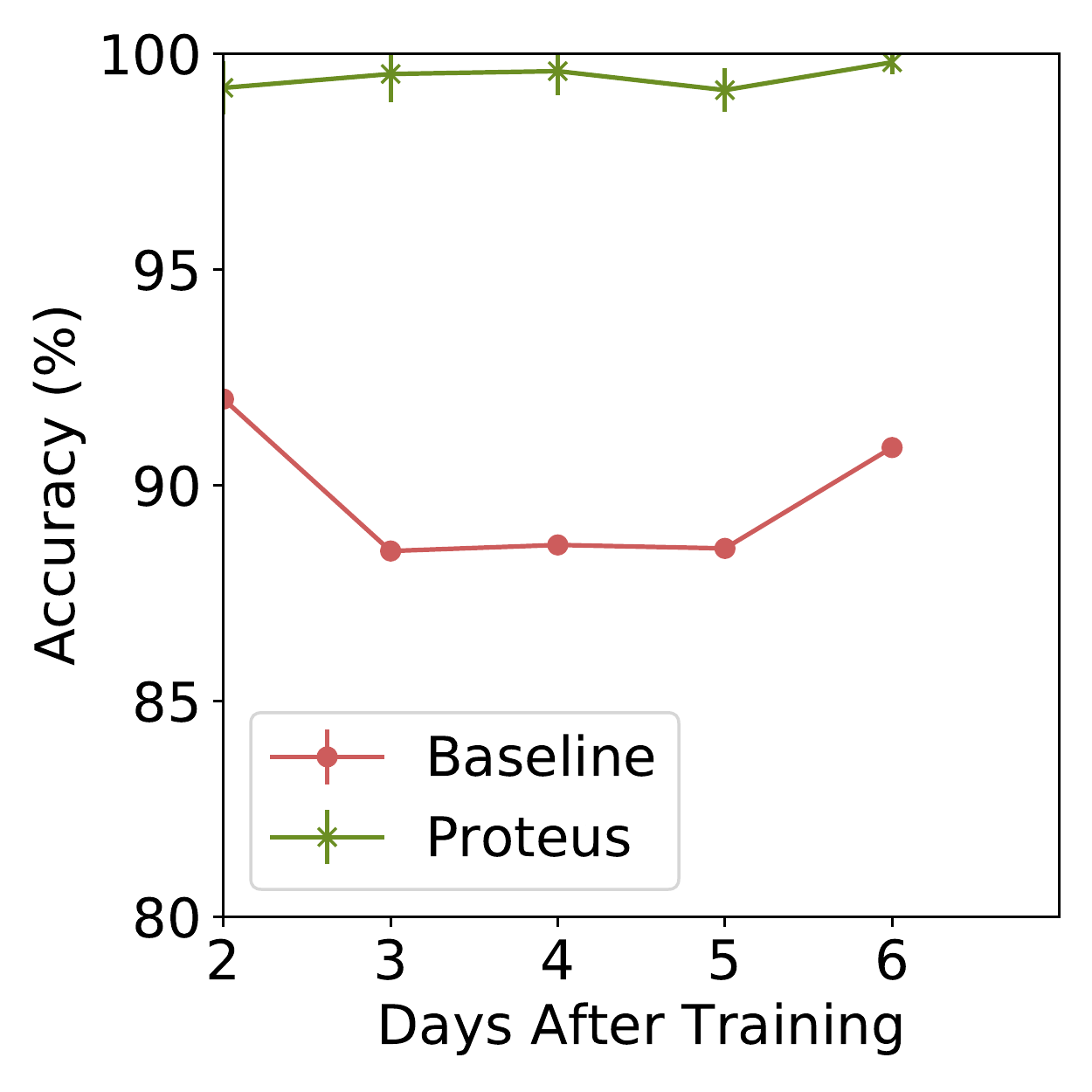}
    \caption{Variation across Time}
    \label{fig:accuracy_time}
    \end{subfigure}
    %\vspace{-0.1in}
    \caption{\name's Accuracy Analysis: \textmd{(a) \name\ achieves high accuracy across indoor (99.8\%), outdoor(95\%), and controlled benchtop experiments (98.2\%). This overshadows the relatively strong auto-correlation baseline by large margins. \name\ continues to achieve high accuracies (~99\%) across (b) different bandwidths, (c) Spreading factors, (d) locations, and (e) time. Classification becomes more challenging for lower BW and higher SF because the separation between lower bandwidths is small and the input only contains a small fraction ($<$1\%) of a chirp for the low bandwidth, high spreading factor configurations. }}
    %\vspace{-0.1in}
    \label{fig:acc_overall}
\end{figure*}

\section{Results}\label{sec:results}
We present the empirical evaluation of \name\ below.

\subsection{Experimental Setup}
To evaluate {\name} we generate a dataset to represent 15 possible classifications for SF ranging from 10-12 and BW of 10.4, 15.6, 125, 250, and 500kHz. Since the preamble of a LoRa packet is a series of upchirps we create a dataset that consists of individual chirps in the form of complex baseband signals, extracted from the preamble of each packet. We use the radio described in Section~\ref{sec:hardware} to transmit LoRa packets and receive using a USRP. With this setup, data was collected in a controlled, indoor, and outdoor setting. 

\begin{itemize}[leftmargin=*, nosep]
    
    \item\textbf{Indoor Data Collection:} We conduct the indoor experiment in an office space. The experiments span six different rooms covering a total area of 1000 sq. ft. The transmitter and receiver are randomly placed in each room. In each setting, we collect data for each class. For each location, we collect data for 800 symbols per class on average. 
    \item\textbf{Outdoor Data Collection:} To emulate an outdoor deployment, we collect data using a campus scale deployment. The gateway is placed at a fixed location on ground level. We move the client either manually or on the top of a car to different locations in the campus area spanning 0.02 sq. mi. For each location, we choose a random SF and a random BW to transmit data. We manually record the GPS coordinates of the location and the configuration used. We collect data for a total of 16 positions on campus. We note that our outdoor data collection is limited, due to enhanced movement restrictions imposed by the COVID-19 outbreak. We will add more outdoor experiments to the set once outdoor experiments become feasible again. 
    \item\textbf{Benchtop Data Collection:} To replicate large distance outdoor experiments, we used a benchtop experiment setup to create a controlled dataset with varying RSSI (receiver signal strength indicator). In this setup, the gateway and client are connected directly over a wire. An attenuator is used to attenuate the transmitted signal using attenuation ranging from 40-140dB for each symbol classification. 
    
Table~\ref{tab:dataset} shows the size of each dataset created from the data collection described above. Each data set has 800 symbols per class and 30 features, with varying number of data points depending on the type of data collection. In addition to data collection described above, for the indoor setting we also collected data over the air on several different days resulting in the Daily dataset. 

\end{itemize}
\begin{table}
\begin{tabularx}{0.45\textwidth} { 
  | >{\centering\arraybackslash}X 
  | >{\centering\arraybackslash}X 
  | >{\centering\arraybackslash}X | }
 \hline
 \textbf{Data Set} & \textbf{Training} & \textbf{Test} \\
 \hline
 Outdoor  & (698,800,30)  & (300,800,30)  \\
 \hline
 Indoor  & (9368,800,30)  & (4016,800,30)  \\
 \hline
 Daily  & (2012,800,30)  & (863,800,30)  \\
 \hline
 Benchtop  & (127754,800,30)  & (54753,800,30)  \\
 \hline
\end{tabularx}
\caption{\textmd{Size of each dataset used by \name}}\label{tab:dataset}
\end{table}

\vspace{6pt}\noindent\textbf{Baseline: } We use a baseline based on the cross-correlation operation. We create an example set that contains one example signal for each class (pair of BW and SF). For a given signal input, $S$, let us say that $f_{S,E_i}(n)$ is the cross-correlation of $S$ with example $E_i$ in class $i$. Then we compute the similarity score for the class $i$ as $score(i) = \max_n{f_{S,E_i}(n)}$.

Finally, the class with the maximum score is assigned to this input. This is a computationally intensive process. Cross-correlation is an $O(N\log(N))$ operation, where $N$ is the length of the signal. We need to perform it for each class.

\subsection{Accuracy Evaluation}
First, we evaluate how accurate the \name\ CNN is in identifying the correct configuration for a packet. Recall, for the \name\ CNN, the raw signal is captured for ~4ms and is used as input for the binary classier. If the received packet is in the low bandwidth category we increase the signal capture to ~65ms, otherwise it remains the same for high bandwidths. This is equivalent to a duration of two chirps (or symbols) for the highest data rate in our experiments (BW=500 kHz and SF=10) and about $\frac{1}{6}$ of a chirp duration for the lowest data rate. The performance of the NN is evaluated by analyzing accuracy for all three scenarios described previously. Our analysis uses a mix of indoor and benchtop data for training the network. 30 percent of the collected data is used for training and all the other data is used for testing. %Fig~\ref{fig:accu} shows an overview of the results.

First, observe Fig.~\ref{fig:accuracy}. \name's CNN achieves a very high overall accuracy of 99.8\%, 95\%, and 98.2\% for indoor, outdoor, and benchtop evaluations respectively. This high accuracy demonstrates the feasibility of \name's core idea, i.e. we can identify the correct configuration for a packet at the gateway with high accuracy. In comparison, the baseline performs significantly worse. For the three settings, the baseline accuracy is 67.5\%, 67\%, and 78\% respectively. One reason for the baseline's worse performance is the challenge of identifying small differences in frequency BW such as 10.4 kHz and 15.6kHz. Unlike the higher BW such as 125 kHz and 250 kHz, these BWs are relatively close and the presence of noise and multipath makes it challenging to differentiate them. 

\vspace{4pt}
\noindent\textbf{Variation across Environments: }Fig.~\ref{fig:accuracy} also demonstrates variation across environments. The system performs better outdoors than indoors. This is mainly because outdoor environments comprise of a lot of free space and relatively less multipath fading. Indoor environments have much more multipath reflections, making them more challenging. For this reason, we envision that the accuracy might be higher in rural outdoor deployments than urban outdoor deployments (since we perform experiments on a campus full of buildings).
Similarly, the accuracy on benchtop data is worse even though it should have low-multipath. Our hypothesis is that this is because the NN is trained on indoor and outdoor data where the signal travels over air, whereas the benchtop experiments are performed over a wire.

\vspace{4pt}
\noindent\textbf{Variation across BW: }Fig.~\ref{fig:accuracy_bw} shows the performance across different BWs. For this experiment we report recall, which is the number of points correctly classified as BW divided by the number of points that were actually transmitted at BW. As shown in the figure, the recall remains around 99\% for all BW with the lowest being 98.4\% (for 15.6 kHz) and the highest being close to 100\% for 10.4 and 125 kHz.

\vspace{4pt}
\noindent\textbf{Variation across SF: }
Fig.~\ref{fig:accuracy_sf} plots the variation of \name's performance over different SFs. As shown in the figure, the recall remains around 99\% for all the three SF values. The recall is slightly lower for the highest SF. This is mainly because the highest SF corresponds to the maximum time for each chirp. This means that if we sample a fixed duration of time, as we do for our inputs, we get the smallest fraction of the chirp for the highest SF. This makes the classification problem more challenging as the SF goes up. Nevertheless, \name\ achieves over 95\% accuracy even at the highest SF used by LoRa by using less than half-percent of a single chirp-duration. This demonstrates the strong performance of \name's CNN design. 

\vspace{4pt}
\noindent\textbf{Variation across Location: }Fig.~\ref{fig:accuracy_location} plots the variation of \name's performance in different physical spaces. L0 to L4 denote four different locations. In each of these locations, \name's accuracy remains consistent being around 99-100\%. 

\vspace{4pt}
\noindent\textbf{Variation across Time: } Fig.~\ref{fig:accuracy_time} plots the variation of \name's performance over time for all 15 classes. In this experiment data was collected over the air for 30 minutes for 5 consecutive days. The accuracy remains almost 100\% for all days. When comparing to the baseline approach, there is a significant drop to ~88\% accuracy.  
\vspace{3pt}

The major takeaway from the accuracy analysis is that \name\ is capable of correctly identifying radio configurations in a diverse set of scenarios with high accuracy. However, a key question we need to answer is, how high of an accuracy is good enough? \name\ achieves an overall accuracy of 97.7\% which translates to less than 1/20 packets lost. This loss becomes quite minimal when considering the overall packet loss for LoRa. In a survey by ~\cite{lpwan_survey}, they show that the packet loss for a BW of 125kHz and a SF of 12 can range from 12\% to 74\% from distances of 0-15km in an outdoor urban scenario. The additional loss that \name\ introduces ends up being a reasonable trade-off.
\begin{figure}[t]
    \centering	
    \includegraphics[width=0.33\textwidth]{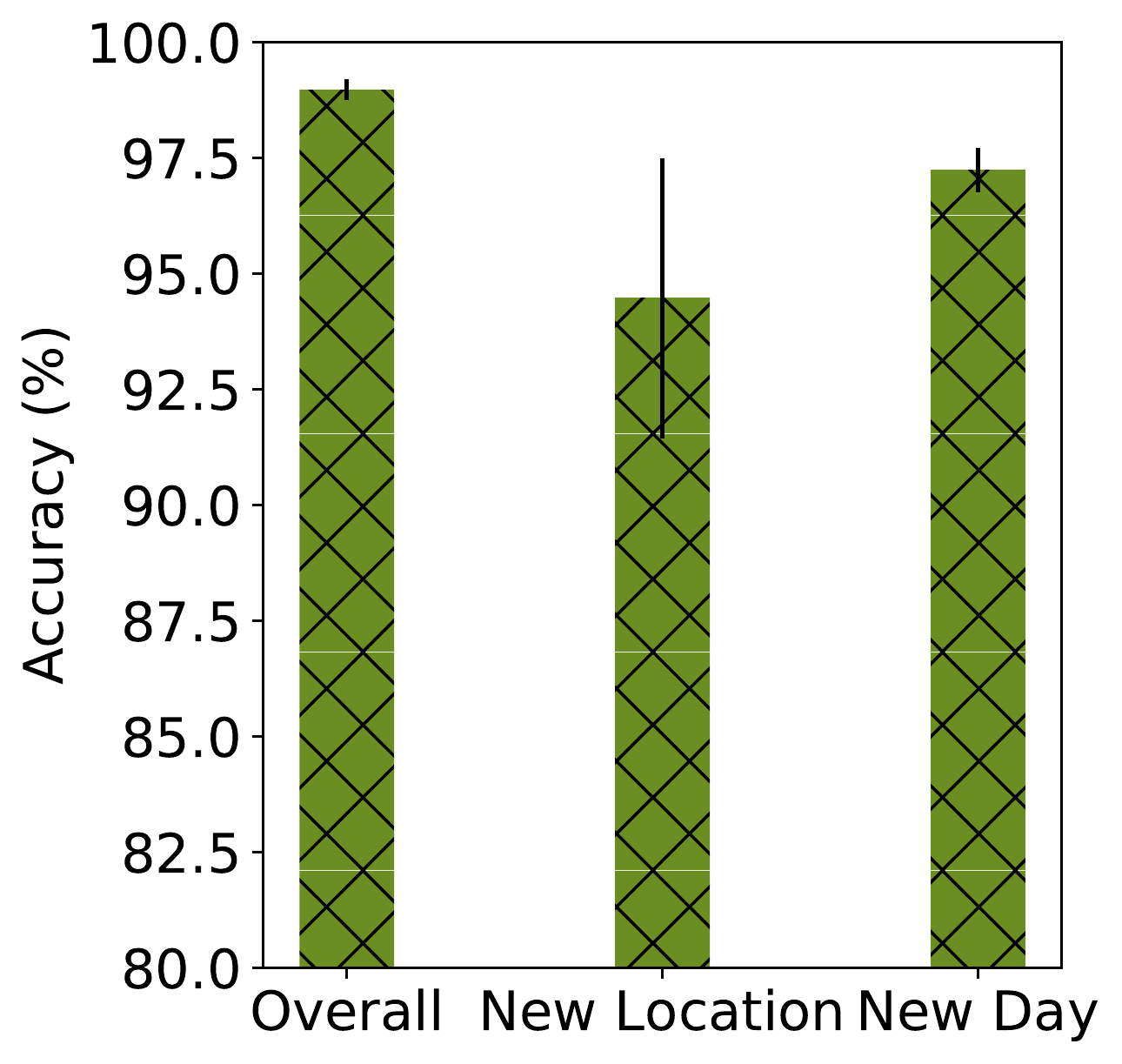}
    \caption{\name\ Generalization: \textmd{When tested in a different location or on a new day, the performance drop is less than 5\%.This shows that \name\ can generalize across new locations and across time. Note that \name\ can be easily fine-tuned in a new location for even better performance.}}
        \label{fig:generalization}

    \vspace{-0.1in}
\end{figure}
\begin{figure}[t]
    \centering
    \includegraphics[width=0.33\textwidth]{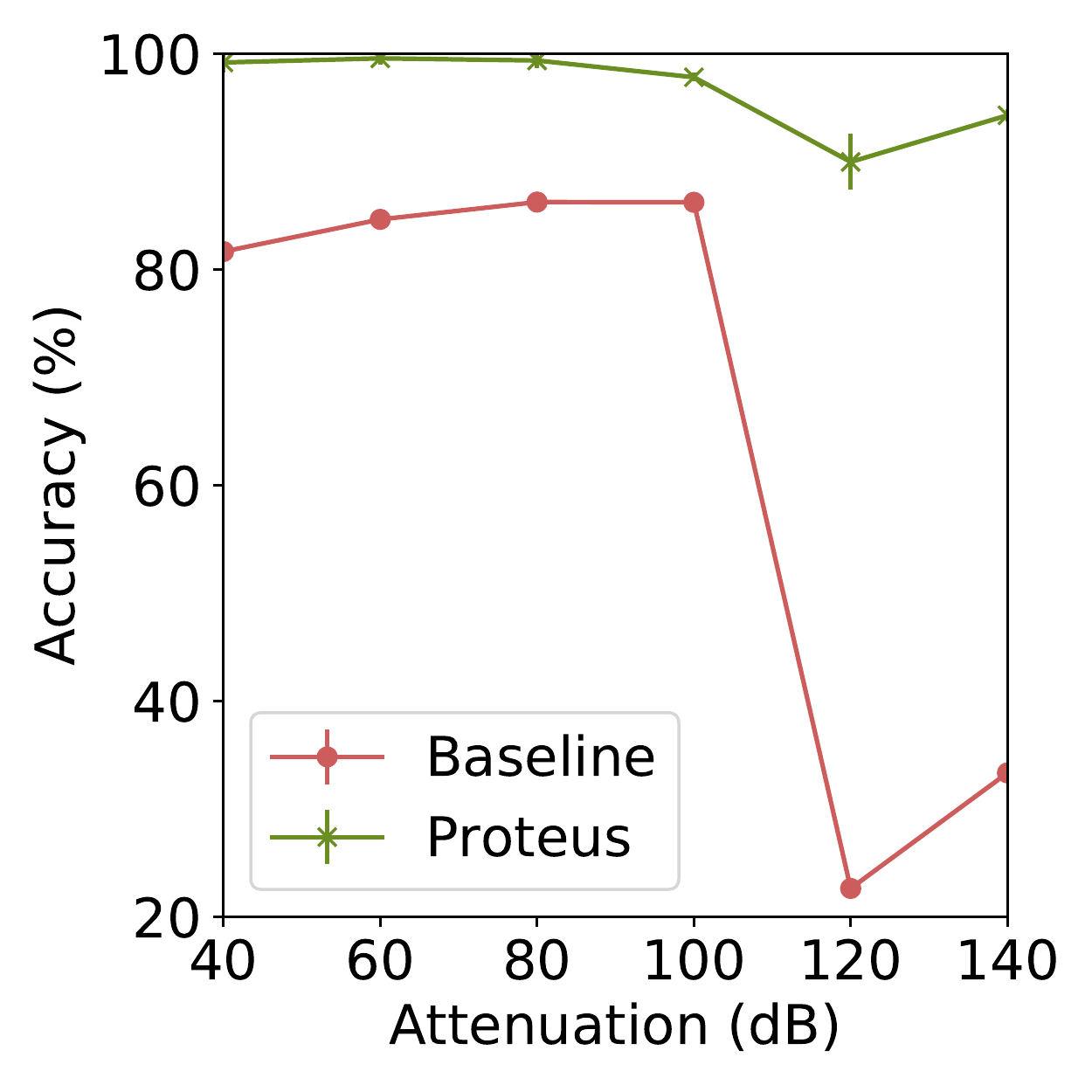}
    \vspace{-0.1in}
    \caption{Sensitivity: \textmd{To determine the sensitivity, we added variable attenuation to the signal path in the benchtop experiment. With increasing attenuation, the performance of the system is sustained, as opposed to the baseline which significantly worsens at lower signal strengths. }}
    \label{fig:attenuation}
    \vspace{-0.1in}
\end{figure}

\subsection{Generalization}
One question that arises for most machine learning frameworks is their ability to generalize to new environments unseen in the training set. We try to answer that question for \name\ using two empirical evaluations.

First, we train the model while excluding two locations from the training data (different rooms in the indoor environment). Specifically, we exclude data obtained from L5 and L6 from the training set. We set apart the data from these two locations for our test set. This allows us to test generalization to new environments. The result of this experiment is plotted in Fig.~\ref{fig:generalization}, which shows the localization accuracy suffers a minor drop from 98.9\% to 94.5\%. 

Second, we test the model for generalization across time. We collect test data on a day that has not been included in the training set (and is set apart by a week). The model maintains the performance it achieved on prior days (97\% accuracy). This shows that there exists some inter-location variation in accuracy, but we didn't observe any temporal variation. The takeaway from this result is that \name\ is able to achieve high accuracy even for input signals of scenarios the CNN has not encountered. This tells us that \name's CNN can be used for a diverse set of LoRa networks.

\subsection{Sensitivity}
LoRa can operate at sensitivity ranging from -149 to -118 dBm depending on the SF and BW settings used. In order for \name\ to be valuable for LoRa network deployments, it also must be able to achieve high accuracy at the same range of sensitivity. To evaluate the accuracy of \name's CNN for signals with low power, we generate a dataset that has been attenuated from 40-140dB and analyze the accuracy of the model. 
Fig.~\ref{fig:attenuation} shows the accuracy of the model as a function of attenuation for \name\ and the baseline method. \name\ has an average accuracy of 96.7\% and maximum of 99\%, regardless of attenuation. This is consistent with accuracy achieved on the overall benchtop experiment reported in Fig.~\ref{fig:accuracy}. On the other hand, the accuracy of the baseline method fluctuates and has a decline for signals exposed to high amounts of attenuation. The overall results tell us that \name's network is robust towards variation in signal strength and should be able to maintain prediction accuracy for the signal conditions LoRa may face.

\begin{figure}[t]
    \centering	
    \begin{subfigure}[b]{0.4\textwidth}
    \includegraphics[width=\textwidth]{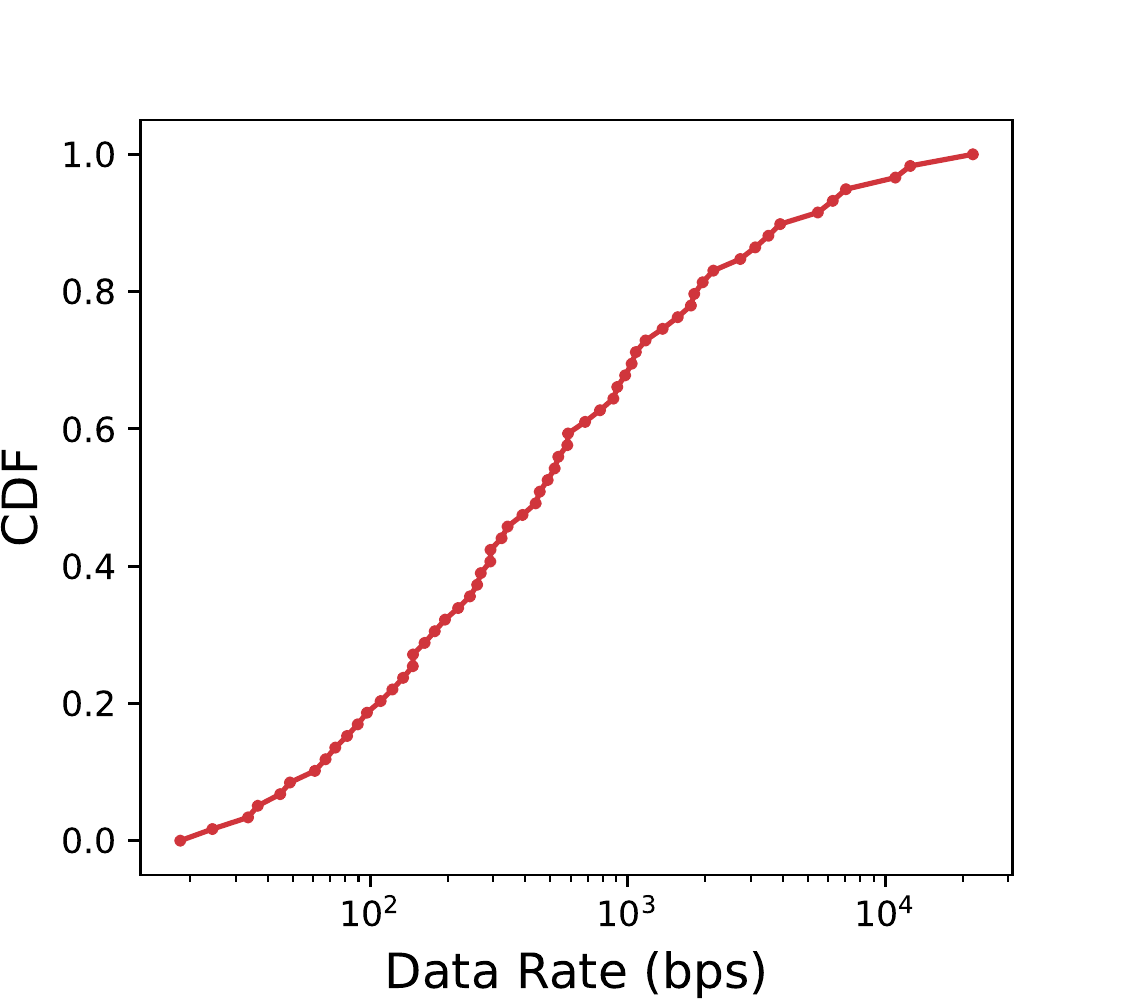}
    \caption{\textmd{Data Rate CDF}}
    \label{fig:cdf}
    \end{subfigure}
    \quad
    \begin{subfigure}[b]{0.4\textwidth}
    \includegraphics[width=\textwidth]{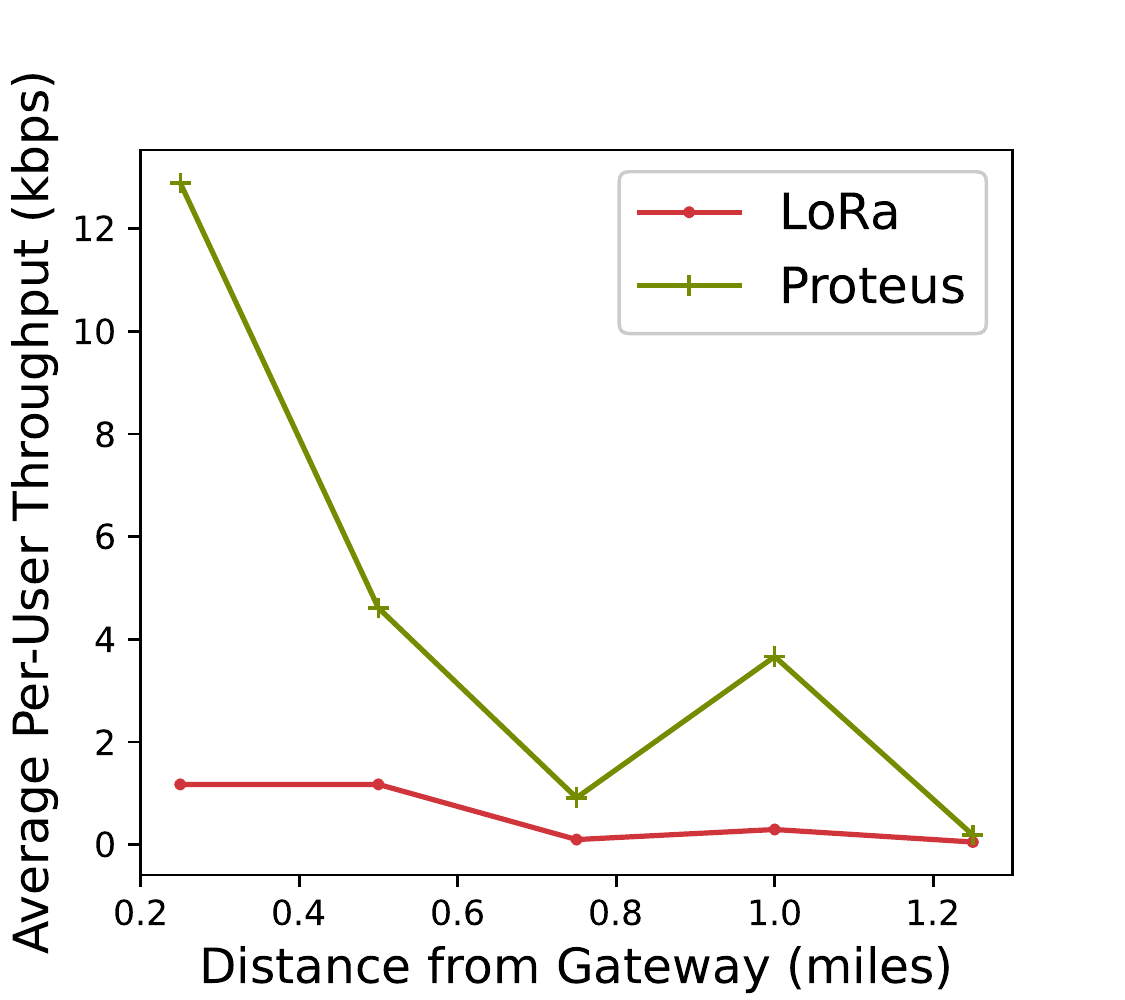}
    \caption{\textmd{Throughput Comparison}}
    \label{fig:th_compare}
    \end{subfigure}
    \caption{Throughput Analysis. \textmd{An analysis of throughput gains provided by \name.}}
    \label{fig:throughput}
\end{figure}

\subsection{Throughput}
To gain a better understanding of the performance increase \name\ can provide, we analyze network throughput. Consider the scenario shown in Fig.~\ref{fig:map}, where a single gateway is deployed to support clients in roughly 0.6 sq. mi.. In a typical LoRa deployment, to support endpoint devices spread across the entire coverage area, we would need to choose encoding parameters that would provide reliable end-to-end communication for \textit{all} devices. In this example, we would be forced to use a BW of 20.8 kHz and SF of 12, which means the maximum achievable per-user throughput would be 48.8 bps (assuming a code rate of 1). On the other hand, \name\ can enable the support of a wide range of encoding parameters. In this particular example it would result in up to 448x increase in achievable throughput per-user in comparison to a standard LoRa deployment. Fig.~\ref{fig:throughput} shows the throughput analysis for \name. We use the scenario in Fig.~\ref{fig:map} and treat each marker as client. In Fig.~\ref{fig:th_compare} we compare the throughput gains of \name\ to a LoRa deployment using a single gateway. The average per-user throughput is calculated per 0.25 mi. region in the coverage map. The results show that \name\ provides higher average user throughput across all regions with a increase ranging from 3.8-11x more than the standard LoRa deployment. 

A key question to answer is, how does the throughput gains of \name\ compare to the Adaptive Data Rate (ADR) mechanism offered by LoRa? ADR is a simple mechanism that can adapt the data rate used by clients in a network by increasing or decreasing the SF based on link budget, along with adjusting transmission power to optimize battery-life~\cite{ADR_info}. However, there are key constraints that come with ADR. First, ADR is not useful in deployments using a single gateway. Since LoRaWAN uses an ALOHA type of protocol, where uplink transmissions can be sent at any time by a client, the gateway would not be able to coordinate switching between different settings and all clients would be required to use the same configurations. For ADR to provide similar throughput gains, multiple gateway deployments would be required. Second, the ADR mechanism requires several control packet exchanges with clients to execute a change in data rate. Executing this exchange can become challenging when RF conditions are unstable and for clients that are mobile or several miles away from the gateway. In ~\cite{ADR_sim}, authors demonstrate through ns-3 simulation that a 100-node network can take from 100-250 minutes for devices to converge to a stable data rate using ADR. Similarly in ~\cite{ADR_Issues}, authors demonstrate that for a 100-node network it takes roughly 200 minutes for a device to switch to a new SF. This convergence time increases even further with larger networks. It is clear that the key drawback with ADR is the number of packet exchanges needed to optimize data rate for any given client. On the other hand with \name, clients can freely update parameters with the trade-off being a potential increase in preamble length. While both solutions can lead to similar improvements in throughput, \name\ would be more advantageous in large-scale networks and single gateway deployments.

\subsection{In the Wild Deployment Feasibility}
In previous sections we demonstrated that \name\ can achieve high accuracy predictions of encoding parameters for LoRa packet transmissions. With this we can enable a dynamic network deployment, resulting in higher overall throughput and support large-scale networks using a single gateway. A final criteria to evaluate is the feasibility of real-world deployments, where the primary constraint is cost. When deploying a wireless network, the primary cost is at the gateway. Not only do we need to account for the gateway hardware but tower infrastructure and maintenance as well, which quickly bring up the cost. For a typical LoRa deployment, to support such a wide range of SF and BW configurations we would need to have multiple gateway deployments. With this in mind, the current design of \name\ is quite reasonable. Nevertheless, the cost of deploying \name\ could be brought down. For instance, instead of using USRP and MacBook Pro, we can scale down to use low-cost small form-factor SDRs and computing platforms.

To justify the use of a low-cost embedded platform we convert our NN model to Tensorflow Lite (TFLite) and deploy on a Raspberry Pi 3 (RPi). The TFLite model is only 275KB in size and can achieve the same accuracy as the original model. While the RPi does not have the same compute capability, we could deploy \name\ on an Up Xtreme instead. The Up Xtreme is a low-power computing platform that provides up to four trillion operations per second and cost $<$\$1000. Finally, a RTL-SDR costs around \$20, can operate at 915MHz, and can have a sampling rate of up to 3.2 Msps, providing the same quality of data as the USRP.

%% file: related.tex
\section{Related Work}
The related work for \name\ falls in three categories: optimization of LPWAN performance, rate adaptation algorithms, and intersection of wireless networks and machine learning (ML).

\vspace{6pt}\noindent\textbf{LPWAN Performance Optimization: }
There has been much past work on improving the performance of LPWANs. The core goals are to maximize throughput and reduce the power consumed by clients while maintaining long-range communication. While we aim to achieve the same goals, \name's axis of improvement is different (and complimentary).

First, one class of research ~\cite{chime, cuomo2018towards,reynders2018improving,cuomo2017explora, abdelfadeel2018fair, bor2017lora} aims to find the ideal combination of transmit parameters for a given network. Specifically, choosing the right combination of frequency and spreading factor can improve network throughput and battery life. \name\ changes this game by allowing a different configuration (BW and SF) for each client device, thereby expanding the scope of past work. By enabling each device to choose its own configuration, \name\ provides additional throughput gains.

Second, some past work has focused on improving performance through new scheduling approaches and power-management systems \cite{abdelfadeel2019free, reynders2018improving, haxhibeqiri2018low}, new protocols to deal with collisions \cite{eletreby_lpwans}, using blind distributed multi-user MIMO to eliminate the need for coordination in receiving concurrent uplink transmissions~\cite{gao2019blind}, or by leveraging multiple gateways to decode weak transmissions \cite{charm}. \name\ differs from this line of work both in terms of techniques and capabilities. \name\ uses a ML approach to identify the configuration used by a client. Using this primitive as the building block, \name\ builds a backwards-compatible system to enable heterogeneous network configuration that can benefit these systems and benefit from this work.

A large part of the development of LPWAN technology has primarily been lead by the industry. In particular, Semtech has developed a variety of LoRa chips that operate in the narrow bands of unlicensed spectrum. Semtech recently released a new chip, the SX1301, a LoRaWAN gateway. This chipset allows clients to operate at multiple SFs without requiring them to inform the gateway a priori. We believe this work supports our hypothesis that dynamic configuration of LoRa clients is the way forward. However, this chip takes a hardware-only approach and is hence limited to different SFs. In contrast, Proteus takes a hardware-software approach and is able to not only identify the correct SF used, but also the BW of the signal being transmitted. Recall that the BW can vary from 7.8kHz to 500 kHz, thus providing a major axis for data rate adaptation.

\vspace{6pt}\noindent\textbf{Machine Learning for Wireless: }There has been a lot of recent work at the intersection of wireless communication and machine learning. For instance, ~\cite{oshea} demonstrates through simulation how a Convolutional Neural Network (CNN) can be used to classify a wide variety of modulation schemes (e.g. BPSK, QPSK, QAM) using complex time domain signals. Similarly, ~\cite{shahid} leveraged CNNs in taking optimal spectrum decisions for LPWAN. With the help of CNN, they minimized interference and enabled classification of LPWAN technologies. In ~\cite{chen2018cognitive}, the authors proposed AI-enabled LPWAN which can select appropriate LPWAN technologies to achieve a better experience from the perspective of traffic control. Our work contributes to and is inspired from this line of work. However, our work differs in terms of the methods used and the overall problem statement. Furthermore, \name\ tackles novel technical challenges in the LPWAN space, i.e. modulation identification from partial symbols, varying time-bandwidth sampling, and real-time operation.

\vspace{6pt}\noindent\textbf{Rate Adaptation: }There is a rich history of rate adaptation algorithms in mobile networks \cite{bicketthesis, gudipati_hotnets, vutukuru_sigcomm, victor_mobicom,camp_mobicom}. As described before, \name's design allows the clients to choose their own rates. We do not propose a rate adaptation algorithm for the client. Instead, we remove the coordination overhead required to convey the rate to the gateway. The closest to our work is \cite{gudipati_hotnets} that proposes a new scheme that does not require coordination. However, \cite{gudipati_hotnets} requires a redesign of the PHY layer, unlike the backwards compatible design offered by \name. 

%% file: conclusion.tex
\section{Discussion and Conclusion}
In this paper we introduce \name, a new gateway design that allows LoRa clients to transmit at their data rate of choice. This enables LoRa gateways to support devices at large-scale over long-range without having to compromise network performance. \name\ uses a CNN to predict the BW and SF of packets transmitted by clients and enable the gateway to decode packets across varying radio configurations. \name\ achieves this by making several key contributions. 

\begin{itemize}[leftmargin=*, nosep]
    \item\textbf{Classifier for LoRa Radio Configurations:} We implement a neural network that can classify 15 different LoRa radio configurations from partial symbols with 99.8\% and  95\%accuracy for indoor and outdoor scenarios.  
    \item\textbf{Real-time Classification:} We demonstrate that radio configurations can be automated and performed in real-time by exploiting the dynamic preamble settings of LoRa.% of LoRa packets. %\name\ relies on up to two preamble symbols to perform classification with high accuracy across a diverse set of scenarios. 
    \item\textbf{Adaptive Sampling:} We implement a multi-stage filter bank to optimize the trade-off between sensitivity, accuracy, and latency of the network. \name\ adapts the BW and capture duration to classify the vast set of radio configuration supported by LoRa. 
\end{itemize}{}

We highlight several future research directions: 
\begin{itemize}[leftmargin=*, nosep]
    \item\textbf{Rate Adaptation.} \name\ can be used to improve rate adaptation techniques for LoRa. Using \name\, clients can configure their own encoding parameters without informing the gateway \textit{a priori}. This avoids much of the typical overhead like control messages, thereby setting the stage for new client-side rate adaptation techniques. %For instance, control messaging between a gateway and client can be minimized. Developing new protocols built upon \name\ for rate adaption has the potential to further boost the performance and efficiency of LPWANs.

    \item\textbf{FPGA Implementation.} While the \name\ neural network performs fast inference, it can be further improved by an FPGA implementation that is integrated into the SDR. Such an implementation can help improve the latency of \name\ by minimizing the time needed to detect, classify, and update radio parameters. 
    
    \item\textbf{Alternative Hardware: } While we develop our system as a gateway augmented with a SDR, off-the-shelf gateways like SX1257\cite{sx1257} already support access to the raw IQ samples of the signal and will be directly compatible with our design.
    
    %\item\textbf{Interpretability} Following recent trends in system design using ML, the \name\ neural network can teach decision trees that are interpretable to enable network operators to debug any misconfigurations.
    %\item\textbf{Network Pruning.} In relation to improving latency, pruning the network used by \name\ is a promising approach. The idea behind network pruning is that with the many parameters in the network, there are bound to be some that are redundant and have insignificant contribution. This minimizes the size of the network and in turn optimizes the time needed to perform classification.
    
    %\item \textbf{Self-adaptive Network.} We can build upon \name to develop a self-adaptive network for LPWANs. A LoRa network  can connect 100s of clients deployed across 
\end{itemize}

As 5G standards are getting finalized, there is increasing interest in defining 6G networks~\cite{6G} -- with a goal of providing an order of magnitude improvement in bandwidth and latency. A promising approach being explored is machine learning, and how devices can auto-configured to communicate with devices, including those using other standards~\cite{6G-ML}. This can significantly reduce control overhead, resulting in increased network capacity. The {\name} architecture is a step towards this vision -- of complete interoperability, while maintaining backwards compatibility with legacy devices. 